\pdfoutput=1 
\documentclass{article}

\usepackage[left=1in, right=1in, top=1in, bottom=1in]{geometry}
\usepackage{amsmath,amssymb,caption}

\usepackage{graphicx, color}
\usepackage{verbatim}
\usepackage{listings}

\usepackage{multicol}
\usepackage{booktabs}
\usepackage{longtable}
\usepackage{multirow}

\usepackage{latexsym}

\usepackage[T1]{fontenc}
\usepackage[utf8]{inputenc}
\usepackage{authblk}

\definecolor{dkgreen}{rgb}{0,0.6,0}
\definecolor{gray}{rgb}{0.5,0.5,0.5}
\definecolor{mauve}{rgb}{0.58,0,0.82}

\usepackage{subcaption}

\usepackage[hyphens]{url}

\usepackage[linkcolor=blue, citecolor=blue, colorlinks=true, urlcolor=blue, breaklinks=true]{hyperref}
\usepackage[numbers,sort&compress]{natbib}
\usepackage{hypernat}
\usepackage[numbers]{natbib}

\usepackage[retainorgcmds]{IEEEtrantools}

\usepackage{nicefrac}

\usepackage[english]{babel}
\usepackage{siunitx}
\DeclareSIUnit\atm{atm}

\def\CC{{C\nolinebreak[4]\hspace{-.05em}\raisebox{.4ex}{\footnotesize ++}}}

\newcommand{ \dydx } [2] { \frac{ \partial #1 }{ \partial #2 } }
\newcommand{ \ddydxx } [2] { \frac{ \partial^2 #1 }{ \partial #2^2 } }

\begin{document}

\title{Recent progress and challenges in exploiting graphics processors in computational fluid dynamics}

\author{Kyle~E.\ Niemeyer\thanks{Kyle.Niemeyer@oregonstate.edu} \thanks{Current address: School of Mechanical, Industrial, and Manufacturing Engineering, Oregon State University}}
\affil{Department of Mechanical and Aerospace Engineering \\
	Case Western Reserve University, Cleveland, OH 44106, USA}


\author{Chih-Jen Sung\thanks{cjsung@engr.uconn.edu}}
\affil{Department of Mechanical Engineering \\
	University of Connecticut, Storrs, CT 06269, USA}

\date{}

\maketitle

\begin{abstract}
The progress made in accelerating simulations of fluid flow using GPUs, and the challenges that remain, are surveyed. The review first provides an introduction to GPU computing and programming, and discusses various considerations for improved performance. Case studies comparing the performance of CPU- and GPU-based solvers for the Laplace and incompressible Navier--Stokes equations are performed in order to demonstrate the potential improvement even with simple codes. Recent efforts to accelerate CFD simulations using GPUs are reviewed for laminar, turbulent, and reactive flow solvers. Also, GPU implementations of the lattice Boltzmann method are reviewed. Finally, recommendations for implementing CFD codes on GPUs are given and remaining challenges are discussed, such as the need to develop new strategies and redesign algorithms to enable GPU acceleration.
\end{abstract}

\section{Introduction}
\label{sec:intro}

The computational demands of scientific computing are increasing at an ever-faster pace as scientists and engineers attempt to model increasingly complex systems and phenomena. This trend is equally true in the case of fluid flow simulations, where accurate models require large numbers of grid points and, therefore, large, expensive computer systems. In particular, high Reynolds numbers---the domain of many real-world flows---require denser computational grids to resolve the scales of importance or even turbulent flows. Modeling reactive flows, with realistic\slash detailed chemical models involving large numbers of species and reactions, can increase the computational demands further by another order of magnitude. Due to the stringent hardware requirements warranted by such computational demands, high-fidelity simulations are typically not accessible to most industrial or academic researchers and designers.

In the past, we could rely on computational capabilities improving with time, simply waiting for the next generation of central processing units (CPUs) to enable previously inaccessible calculations. However, in recent years, the pace of increasing processor speeds slowed, largely due to limitations in power consumption and heat dissipation preventing further decreases in transistor size. Power consumption, and, therefore, heat output, scales with processor area, while processor speed scales with the square root of area. Adding cores allows processors to increase overall performance via parallelism while minimizing larger power consumption by avoiding significant increases in overall area. In order to keep up with Moore's law---processor speeds doubling roughly every year and a half---processor manufacturers are embracing parallelism. Top-of-the-line CPUs used in personal computers and supercomputing clusters contain four to eight cores. Graphics processing units (GPUs), on the other hand, consist of many hundreds to thousands of---albeit fairly simple---processing cores, and fall in the category of ``many-core'' processors. This level of parallelism matches that of large clusters of CPUs.

Regarding the need for parallel computing, it would depend on whether the types of problems can be parallelized or whether they are fundamentally serial calculations. The key to parallel problems (sometimes termed ``embarrassingly parallel'') is data independence: multiple tasks must be able to operate on data independently. Examples of embarrassingly parallel problems include matrix multiplication, Monte Carlo simulations, calculating cell fluxes across space in the finite volume approach, and calculating finite differences across a grid. In contrast, calculating the Fibonacci sequence is inherently serial, as each term relies on the previous two. In general, iterative solution methods are serial, although the internal calculations of each iteration might be parallelizable.

As researchers identify computing problems with appropriate data parallelism, GPUs are becoming popular in many scientific computing areas such as molecular dynamics~\cite{Elsen:2006gc,Stone:2007hk,Anderson:2008bt,Friedrichs:2009ir,Hardy:2009kk}, protein folding~\cite{Beberg:2009jw}, quantum chemistry~\cite{Vogt:2008gk,Yasuda:2008bb,OlivaresAmaya:2010bc}, computational finance~\cite{Surkov:2010eh,Pages:2011cw,Fatone:2012cd}, data mining~\cite{Jian:2011fm}, and a variety of computational medical techniques such as computed tomography scan processing~\cite{Sherbondy:2003kt,Mueller:2006wq,Birk:2011fr,Birk:2012fx}, white blood cell detection and tracking~\cite{Boyer:2009ct}, cardiac simulation~\cite{Nimmagadda:2011in}, and radiation therapy dose calculation~\cite{Pratx:2011fm}. In 2007 and 2008, Owens et al.~\cite{Owens:2007ts,Owens:2008ku} surveyed the use of GPUs in general purpose computations, but as the popularity of GPU acceleration exploded in recent years many more areas are under investigation. Most efforts moving existing applications to GPUs demonstrated around an order of magnitude (or more, in some cases) improvement in performance.

This survey is structured as follows. First, we introduce GPU computing and discuss topics related to programming and optimizing the performance of applications running on graphics processors. Second, we present two case studies relevant to computational fluid dynamics (CFD): a finite-volume Laplace equation solver for heat conduction in a square plate, and a finite-difference incompressible Navier--Stokes solver for lid-driven cavity flow. We use these simple examples to demonstrate the potential performance improvement of GPU codes relative to CPU versions, and also discuss various methods to optimize GPU applications. Next, we review efforts to both modify existing and create new CFD solvers for GPU acceleration, covering laminar, turbulent, and chemically reactive flow solvers. In addition, we briefly review efforts to accelerate the relatively new lattice Boltzmann method (LBM), for simulating fluid flows, on GPUs. Finally, we discuss the progress made thus far to exploit GPUs for CFD simulations as well as remaining challenges, and make some recommendations.

Note that while the current work focused on GPUs, the developed approaches can apply to future many-core processing architectures in general. In fact, in the case of the OpenCL programming language~\cite{Munshi:2011wk}, the same programs written for GPUs now should be useable on whatever many-core processing standard is adopted---it is designed for executing programs on heterogeneous computing platforms. There is a trend toward massively parallel processors, and GPUs are an early entry in this category.

\section{GPU Computing}
\label{sec:gpu}

\begin{figure}[tbp]
\begin{center}
\includegraphics[width=.8\linewidth]{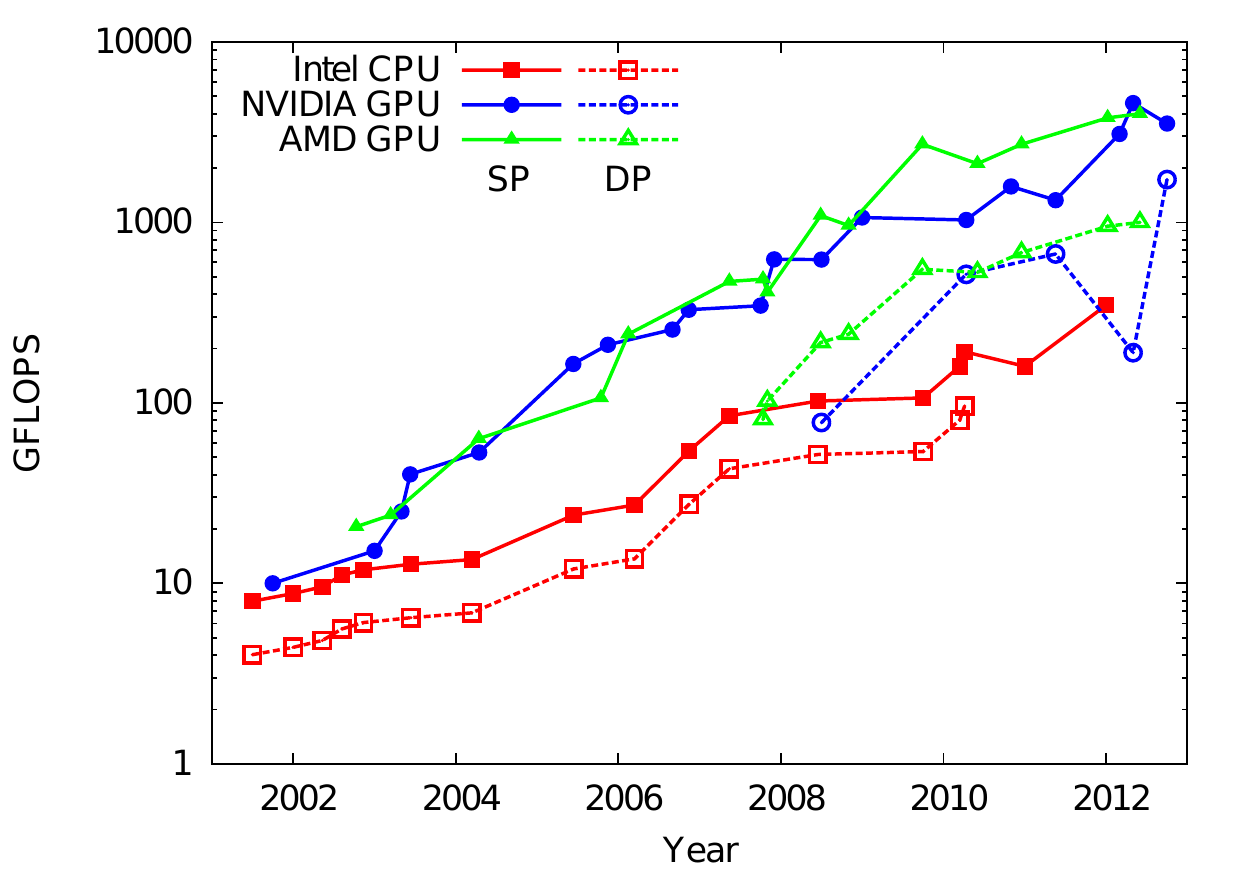}
\caption{Theoretical peak performance of Intel CPUs and NVIDIA\slash AMD GPUs, measured in GFLOPS (gigaFLOPS, i.e., billion floating-point operations per second), over the past decade. ``SP'' and ``DP'' refer to single and double precision, respectively. Note that the performance is presented in logarithmic scale. Source: John Owens personal communication and vendor specifications.}
\label{fig:flops}
\end{center}
\end{figure}

As their name suggests, GPUs were initially developed for graphics\slash video processing and display purposes, and programmed with specialized graphics languages. In these applications, where many thousands to millions of pixels need to be displayed onscreen simultaneously, throughput is more important than latency. In contrast, CPUs execute a single instruction (or few instructions, in the case of multiple cores) rapidly. This led to the current highly parallel architecture of modern graphics processors. The explosive growth of GPU processing capabilities---as well as diminishing costs in recent years---has been propelled mainly by the video game industry's demands for fast, high-quality processing (both for onscreen images and physics engines), and these trends will likely continue driven by commercial demand. Figure~\ref{fig:flops} demonstrates this recent growth in performance, comparing the theoretical peak floating-point operations per second (FLOPS) of modern multicore CPUs and GPUs; the newest GPUs offer more than an order of magnitude higher performance, although no data was available for the most recent CPU models.

\subsection{Programming GPUs}

The current generation of GPU application programming interfaces (APIs), such as CUDA~\cite{NVIDIA:2011wf} and OpenCL~\cite{Munshi:2011wk}, enables a C-like programming experience while exposing the underlying massively parallel architecture. Fortunately, this avoids programming in the graphics pipeline directly. We will focus our discussion on CUDA, a programming platform created and supported by NVIDIA, but OpenCL, an open-source framework supported by multiple vendors, is similar so the same concepts apply (albeit with slightly different names for equivalent features). This section is not intended to function as a complete reference for programming CUDA applications, but only to give a brief overview of the CUDA paradigm. Interested readers should see the textbooks, e.g., by Kirk and Hwu~\cite{Kirk:2010we} and Sanders and Kandrot~\cite{Sanders:2010tq}.

In CUDA, a parallel function is known as a ``kernel,'' which consists of many threads that perform tasks concurrently. Functions intended for operation on the GPU (the device) and the CPU (the host) are preceded with \texttt{\_\_device\_\_} and \texttt{\_\_host\_\_}, respectively. Kernel functions are indicated with \texttt{\_\_global\_\_}. Threads are organized into three-dimensional blocks, which in turn are organized into a two-dimensional grid.\footnote{Recent GPU hardware allows a three-dimensional grid.} All threads in a grid execute the same kernel function. The specific location (coordinate) of a thread inside the hierarchy of blocks and grids can be accessed using the variables \texttt{threadIdx} and \texttt{blockIdx}; the dimensions of the block (i.e., the number of threads) and grid (i.e., the number of blocks) can be retrieved using \texttt{blockDim} and \texttt{gridDim}, respectively.

Figure~\ref{fig:vector_add} shows a simple kernel function for adding two vectors, compared with an equivalent CPU function. Note that instead of looping through the elements, the threads of the kernel function independently and concurrently add the elements of the two vectors. In general, parallelizing applications for use on GPUs follows this pattern, replacing loops with kernel functions where data may be operated on independently. In this example, both the addend vectors and the sum vector are stored in the GPU's global memory, which is accessible to all threads in a kernel. Memory on the device must be allocated using the \texttt{cudaMalloc} function prior to the kernel launch, and memory must be explicitly transferred between the host and the device outside the kernel using the \texttt{cudaMemcpy} function. The structures of the thread blocks and grid are specified using the \texttt{dimBlock} and \texttt{dimGrid} variables. In the example given in Fig.~\ref{fig:vector_add}, for simplicity, both are one-dimensional arrays, with the grid consisting of one block for each element in the vector.

\begin{figure}[tbp]
\begin{center}
    \begin{lstlisting}
// C vector addition
void vector_Add (int size, const float *a, 
    const float *b, float *c) {
  for (int i = 0; i < size; ++i) {
    c[i] = a[i] + b[i];
  }
}
int main (void) {
  int size = 10;
  ...
  vector_Add (size, A, B, C);
  ...
}
    \end{lstlisting}
  \end{center}
\begin{center}
    \begin{lstlisting}
// CUDA vector addition
__global__ void vector_Add_CUDA (int size, 
    const float *a, const float *b, float *c) {
  int ind = blockIdx.x * blockDim.x + threadIdx.x;
  c[ind] = a[ind] + b[ind];
}
int main (void) {
  int size = 10;
  ...
  dim3 dimGrid (size, 1); dim3 dimBlock (1, 1);
  vector_Add_CUDA <<<dimGrid, dimBlock>>>(size, Ad, Bd, Cd);
  ...
}
    \end{lstlisting}
  \end{center}
  \caption{Examples of vector addition on the CPU (top) and the GPU (bottom).}
  \label{fig:vector_add}
\end{figure}

Another avenue for accelerating applications using GPUs is OpenACC~\cite{OpenACC:2011vn,Reyes:2012wg}, which uses compiler directives (e.g., \texttt{\#pragma})  placed in Fortran, C, and \CC\ codes to identify sections of code to be run in parallel on GPUs. This approach is similar to OpenMP~\cite{Dagum:1998hb,Chandra:2001ts,Board:2011wl} for parallelizing work across multiple CPUs or CPU cores that share memory. OpenACC is an open standard being jointly developed by NVIDIA, Cray, the Portland Group, and CAPS.  Since OpenACC is relatively new and immature, only a few groups have used OpenACC thus far to accelerate their applications. Wienke et al.~\cite{Wienke:2012vj} found that OpenACC achieved 80\% of the performance of OpenCL in simulations of bevel gear cutting, but only 40\% in solving the neuromagnetic inverse problem in the neuroimaging technique magnetoencephalography (reconstructing focal activity in the brain). Reyes et al.~\cite{Reyes:2012} showed a similar range of performance, comparing OpenACC with CUDA implementations of LU decomposition, a thermal simulation tool, and a nonlinear global optimization algorithm for DNA sequence alignments. Recently, Levesque et al.~\cite{Levesque:2012} reported on their experience hybridizing Sandia National Laboratory's massively parallel direct numerical simulation code S3D from MPI-only to MPI\slash OpenMP\slash OpenACC for three levels of parallelism; we will discuss their results in greater detail in Section~\ref{sec:reactive}.

\begin{figure}[tbp]
\begin{center}
    \begin{lstlisting}
// OpenACC vector addition
#pragma acc kernels
void vector_Add (int size, const float *restrict a, 
    const float *restrict b, float *restrict c) {
  for (int i = 0; i < size; ++i) {
    c[i] = a[i] + b[i];
  }
}
int main (void) {
  int size = 10;
  ...
  vector_Add (size, A, B, C);
  ...
}
    \end{lstlisting}
  \end{center}
  \caption{Example of vector addition using OpenACC directives.}
  \label{fig:vector_add_acc}
\end{figure}

Figure~\ref{fig:vector_add_acc} shows the vector addition example with OpenACC compiler directives. With the exception of the \texttt{restrict} keyword added to the function arguments, the only modification to the original CPU version is the single \texttt{\#pragma acc} line added before the loop. The main benefit of the OpenACC approach (as well as OpenMP) is that compatible programs may be accelerated without modifying the underlying source code---a non-OpenACC-enabled compiler would treat the directives as comments. Fortran code is handled similarly, albeit with a different directive indicator syntax (\texttt{C\$ACC} or \texttt{!\$acc} rather than \texttt{\#pragma acc}). This contrasts greatly with porting applications written for the CPU to either CUDA or OpenCL, which must be completely rewritten. This convenience comes at the cost of slightly degraded performance, but OpenACC allows researchers to accelerate existing code in a matter of hours, rather than days or weeks. We will compare the performance of OpenACC implementations of our case studies in Section~\ref{sec:case}.

\subsection{GPU performance considerations}

In this section we will discuss some topics related to GPU performance, with an emphasis on configuring appropriate device memory and thread execution. For a more comprehensive source, see the textbook by Kirk and Hwu~\cite{Kirk:2010we}. As before, we focus on CUDA programming and its naming conventions, while the same principles apply to OpenCL.

Selecting appropriate memory types for different data is the first place to begin improving the performance of a GPU program. Global memory, which the CPU uses to transfer data to and from the GPU, is accessible to all threads in a grid. However, accessing the global memory is fairly slow, and many threads attempting to access the global memory will build up traffic congestion---further slowing communication. In fact, one measure of the performance of a GPU application is the compute to global memory access (CGMA) ratio, which is the ratio of floating point computations to global memory access calls. If the CGMA ratio is around one, then the performance of a GPU application will be limited by the global memory access latency rather than the floating-point processing speed of the particular GPU hardware. A GPU application can only achieve best performance if the CGMA ratio is much higher than one. Typically, excess global memory use is eliminated by using other, faster GPU memory types.

Constant memory offers one alternative to global memory when global access is needed. This is read only, and offers high bandwidth when all threads access the same memory location simultaneously. The CPU transfers data to constant memory on the device before a kernel is executed, and this data cannot be modified by the GPU. Similar to constant memory is texture memory, which is also read-only and available to all threads. Texture memory is cache-optimized for two-dimensional access, as it is a descendent of the GPU's display capabilities (textures map a two-dimensional image to a three-dimensional surface).

There are also device memory types accessible at the block and thread levels. Shared memory is allocated for each thread block, and is an efficient way for threads in the same block to cooperate---it is roughly 100 times faster than global memory. Registers are private memory blocks available to each thread which also offer fast access. In addition, threads have access to private local memory, which is actually stored in the global memory (and has the corresponding slow access time). Automatic arrays (arrays declared with non-constant size) are stored in the local memory in CUDA, so these should be avoided. Instead, only arrays with a constant size when compiling should be used. Since each GPU offers a limited amount of memory, properly configuring memory is an important task in designing an application. In general, memory will be the limiting factor governing the number of concurrent threads; for example, each thread block offers a limited amount of shared memory and registers.

Another performance consideration relates to the execution of threads. Recall that threads are organized into blocks, which are in turn organized in a grid. Thread blocks can be executed by the GPU in any order, but blocks are not necessarily execution units themselves. Instead, blocks are partitioned into ``warps'' for execution. In the current generation of CUDA devices, each warp consists of 32 threads. If a block consists of more than 32 threads, the block is partitioned into multiple warps based on the thread index (e.g., \texttt{threadIdx}). A block whose size is not divisible by 32 will be padded with extra threads. All threads in a warp must follow the same instruction path, otherwise threads will diverge and reduce performance significantly. For example, if some threads in warp execute the \texttt{if} statement in an \texttt{if-then-else} construct, while others follow the \texttt{else} path, the GPU can no longer execute the threads concurrently and multiple passes are required (in this example, doubling the execution time). To avoid thread divergence, thread blocks should be organized so that warps follow the same control paths.

It is impossible to avoid using global memory since it is the primary route to transferring data between the CPU and GPU. One way to improve the performance of global memory access is to exploit ``memory coalescing'' techniques. Understanding coalescing requires some insight into the physical nature of global memory. On CUDA-enabled GPU devices, global memory is typically implemented using dynamic random access memory (DRAM), the same type used on personal computers and workstations. DRAM stores bits of data as tiny electrical charges in small capacitors; reading memory from DRAM cells requires a sensor to share and measure these charges. In order to speed up this relatively slow procedure, the sensor accesses consecutive memory locations around the requested location to increase the data read rate. This hardware behavior can be exploited by instructing threads in the same warp to access consecutive memory locations. If this is detected, the GPU will automatically coalesce (or combine) these memory accesses into a single operation, allowing much higher global memory bandwidth. Interested readers should see Kirk and Hwu~\cite{Kirk:2010we} and Jang et al.~\cite{Jang:2011ct} for more detail and examples.

Current-generation GPUs have limited bandwidth to process instructions (e.g., floating-point calculations, conditional branches). One common way to improve performance by removing unnecessary instructions is to perform loop unrolling. This avoids both conditional branch instructions (checking if the loop is finished) and the loop counter update. Also, the indices of accessed arrays are now constants rather than changing variables, enabling further optimization. In some compilers, this can be achieved with a \texttt{\#pragma unroll} compiler directive preceding the loop, but, where practical, manual unrolling ensures high performance.

Another consideration that is particularly relevant to scientific computations on the GPU is the use of hardware-accelerated transcendental functions, which are significantly faster than corresponding software versions. These can be called by prefixing functions with ``\texttt{\_\_}'', e.g., \texttt{cos()} becomes \texttt{\_\_cos()}. Currently, these hardware functions are limited to single-precision calculations; only the software versions of double-precision functions are available, although this may change in the future. The enhanced performance comes at the cost of slightly reduced accuracy. For example, the maximum ulp (``units in the last place'') errors of the software \texttt{exp(x)} and hardware \texttt{\_\_exp(x)} are 2 and $2 + \text{floor} \left(\left|1.16x\right| \right)$, respectively. The CUDA Programming Guide describes the error of all the available hardware and software functions~\cite{NVIDIA:2011wf}. Automatic use of the hardware functions can also be achieved with the compiler flag ``\texttt{-use\_fast\_math},'' which automatically converts all potential (single-precision) functions to their hardware equivalents.

\section{Case studies}
\label{sec:case}

We performed two case studies, relevant to CFD, in order to demonstrate the potential acceleration of CFD applications using graphics processors. For both studies, four versions were compared: single-core CPU, six-core CPU using OpenMP, native GPU using CUDA, and GPU-accelerated using OpenACC. The GPU performance experiments were performed using an NVIDIA Tesla c2075 GPU with \SI{6}{\giga\byte} of global memory. An Intel Xeon X5650 CPU, running at \SI{2.67}{\giga\hertz} with \SI{256}{\kilo\byte} of L2 cache memory per core and \SI{12}{\mega\byte} of L3 cache memory, served as the host processor for the GPU calculations and ran the single-core CPU and OpenMP calculations.

We used the GNU Compiler Collection (gcc) version 4.6.2 (with the compiler options ``\texttt{-O3 -ffast-math -std=c99 -m64}'') to compile the CPU programs, the PGI Compiler toolkit version 12.9 to compile the OpenMP (``\texttt{-fast -mp}'') and OpenACC (``\texttt{-acc -ta=nvidia,cuda4.2,cc20 -lpgacc}'') versions, and the CUDA 5.0 compiler nvcc version 0.2.1221 (``\texttt{-O3 -arch=sm\_20 -m64}'') to compile the GPU version. The functions \texttt{cudaSetDevice()} and \texttt{acc\_init()} were used to hide any device initialization delay in the CUDA and OpenACC implementations, respectively.

\subsection{Laplace solver}
\label{sec:laplace}

\subsubsection{Methodology}
\label{sec:laplace_meth}

The first case study we performed consisted of solving Laplace's equation for heat conduction in a square plate. The boundary conditions were a constant zero (nondimensionalized) temperature along the sides and bottom, and a constant temperature of one along the top. Laplace's equation alone is a fairly trivial example, but due to its relevance to many approaches to solving the pressure term in the Navier--Stokes equations we included it here. Using the finite volume method with a constant grid, the discretization of the equation is:
\begin{align}
\nabla^2 T &= \ddydxx{T}{x} + \ddydxx{T}{y} = 0 \\
a_P T_P &= a_W T_W + a_E T_E + a_S T_S + a_N T_N + S_u \label{eq:temp} \\
a_P &= a_W + a_E + a_S + a_N - S_P \\
a_W &= a_E = k \frac{\delta y}{\delta x} \\
a_S &= a_N = k \frac{\delta x}{\delta y}
\end{align}
where \emph{k} is the thermal conductivity, $\delta x$ and $\delta y$ are the grid spacing in the $x$ and $y$ directions respectively, and $S_u$ and $S_P$ represent source terms used for boundary conditions. All quantities are dimensionless. For the given boundary conditions,
\begin{align}
x &= 0 : \quad a_W = 0 \\
x &= 1 : \quad a_E = 0 \\
y &= 0 : \quad a_S = 0 \\
y &= 1 : \quad \begin{cases}
	a_N = 0 \\
	S_u = 2 k w \frac{\delta x}{\delta y} \\
	S_P = -2 k w \frac{\delta x}{\delta y} \end{cases}
\end{align}
where $w$ is the thickness of the plate.

We solved Eq.~\eqref{eq:temp} iteratively using the red-black Gauss--Seidel (GS) method with successive over-relaxation (SOR). Traditional GS or SOR approaches are not suitable for use on parallel CPU or GPU systems since the order of operations is neither known nor controllable, and conflicts in accessing and writing to memory may occur (although Jacobi iteration is suitable for parallel\slash GPU implementation, since calculation of new values depends only on old values). Red-black SOR solves this problem by coloring the grid like a checkerboard, alternating red and black cells. First, the algorithm updates values at red cells---which depend only on black cells---then black cells---which depend only on red cells.  Both of these operations can be performed in parallel. Red-black SOR was first used to solve a system of linear equations on vector and parallel computer systems by Adams and Ortega~\cite{Adams:1982}, although introduced earlier (e.g., by Young~\cite{Young:1971}). Liu et al.~\cite{Liu:2011} provided a more detailed analysis of red-black SOR implemented on GPUs.

In order to show the importance of redesigning algorithms for GPUs, we enabled flags in the code for various optimization steps. The initial, naively implemented GPU code matched the original serial CPU version: two GPU kernel functions updated the temperatures in the red and black cells, returning residual values for every cell each iteration (in order to determine when the SOR algorithm stops). In a first optimization step, we organized the thread blocks such that threads in the same warp access adjacent locations in global memory to activate coalescing. Next, we improved this coalesced memory access by storing the temperature values for the red and black cells in two arrays---such that read and write operations always accessed neighboring memory locations, rather than every other. Finally, to minimize the GPU-CPU memory transfer each iteration, we used shared memory to calculate the maximum residual of each block, such that only a single value per block needed to be transferred back to the CPU rather than values for all threads. Additionally, in order to avoid thread divergence caused by conditional statements for boundary conditions, ``ghost cells'' that held constant temperatures of zero surrounded the computational domain. Therefore, at the boundaries these cells could be accessed instead of needing a conditional statement to avoid an out-of-bounds array access error.

We also explored using texture memory to store the constant coefficient arrays (e.g., $a_P$, $a_W$, etc.), but found the performance to be equivalent or worse than coalesced global memory. In addition, for single-precision calculations ``atomic'' memory operations can be used to allow threads in different blocks to access the same points in global memory. This enables global reduction operations, in this case allowing a single residual value to be transferred from the GPU to the CPU per iteration rather than one per block as with the shared memory alone. We found that using atomic operations improved performance about 5\% or less; the savings in memory transfer likely balanced the generally low performance of such operations.

The OpenACC solver was based on the CPU solver, with directives instructing the compiler to use the GPU on loops matching the kernel functions of the GPU solver---no other changes to the underlying CPU code were made. The OpenMP solver was created the same way, using the appropriate compiler directives. No specific optimization instructions were given to either the OpenMP or OpenACC solver; rather, we allowed the compiler to manage this automatically.

In order to study the performance of the native GPU and OpenACC solvers against the CPU versions, we varied the mesh size from 128\textsuperscript{2} to 8192\textsuperscript{2}. The source code was written in standard and CUDA C for the CPU and GPU versions, respectively, with compiler directives added to the CPU version to create OpenMP and OpenACC versions.\footnote{The full source code is available: \url{http://github.com/kyleniemeyer/laplace_gpu}} In the fully optimized (corresponding to the shared-memory implementation) CUDA-based GPU version, we kept a constant block size of 128 $\times$ 1 (except for the case of the mesh size being 128, where we used 64 $\times$ 1), aligned with either the vertical or horizontal directions depending on if the global memory coalescing flag was enabled (arranged so that thread warps aligned with adjacent locations in memory, as described above). The naive, coalesced global memory, and improved coalescing configurations used grid sizes of $N \times \nicefrac{N}{B}$, $\nicefrac{N}{B} \times N$, and $\nicefrac{N}{2 B} \times N$, respectively, where \emph{N} is the number of mesh cells in one direction (e.g., 512) and \emph{B} is the block size (e.g., 128). The fully-optimized, shared memory configuration built on the improved coalescing configuration and thus used the same grid size. In order to perform a fair comparison, the serial CPU code---and, therefore, the OpenMP and OpenACC versions---used the red-black SOR algorithm, with separate arrays for the red and black pressure values in the same manner as the optimized GPU versions.

\subsubsection{Results}

\begin{figure}[tbp]
\begin{center}
\includegraphics[width=.8\linewidth]{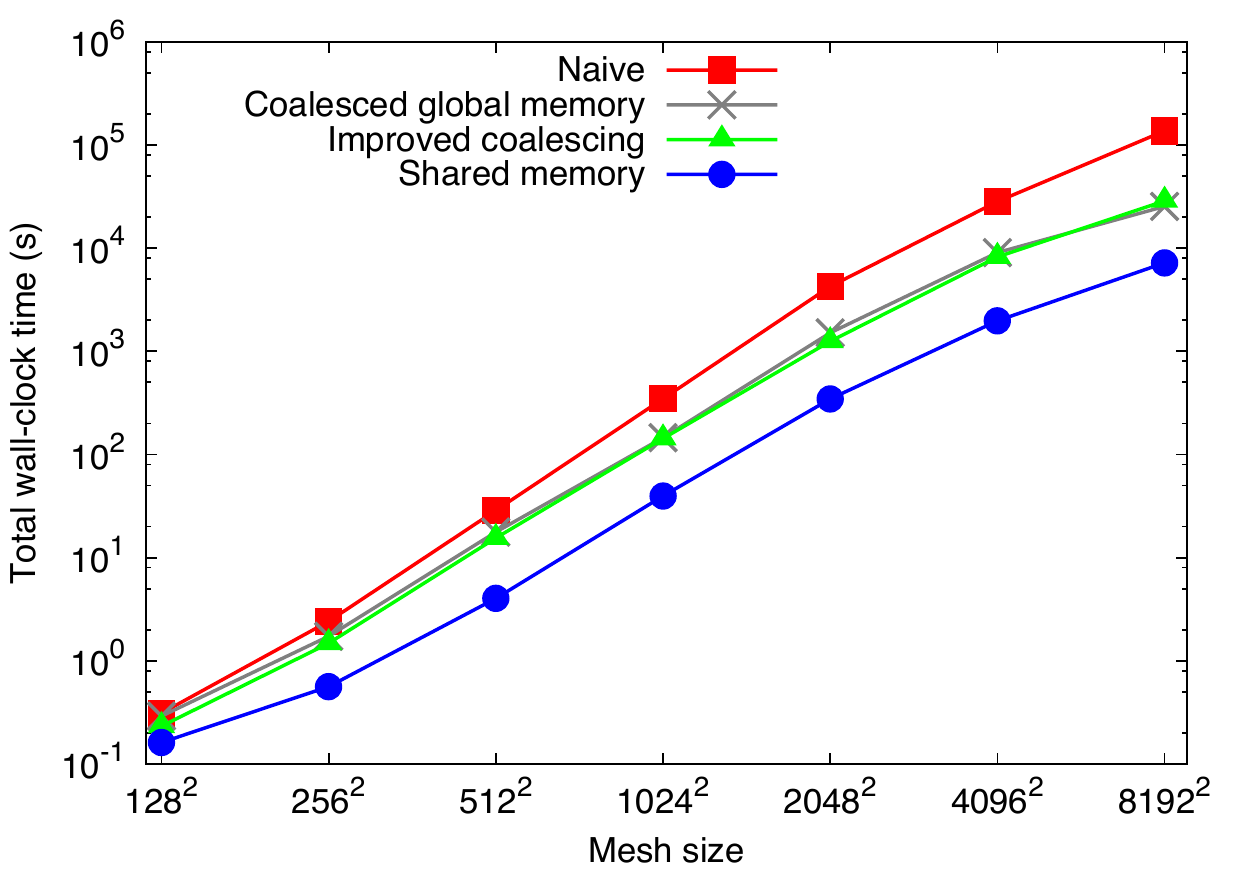}
\caption{Performance of GPU solver with various optimization approaches, for a wide range of mesh sizes. The ``shared memory'' implementation is fully optimized.}
\label{fig:laplace_gpu_opt}
\end{center}
\end{figure}

First, we demonstrate the importance of redesigning algorithms for GPU computing, taking into consideration GPU-specific performance improvements. Figure~\ref{fig:laplace_gpu_opt} shows the performance of the naive GPU solver and with various optimization steps for double-precision calculations. Single-precision results showed similar trends, requiring about half the computational time. By specifically optimizing the code for execution on graphics cards, we increased the performance of the GPU code by up to a factor of nearly 19 compared to the naive implementation.

\begin{figure}[tbp]
\begin{center}
\includegraphics[width=.8\linewidth]{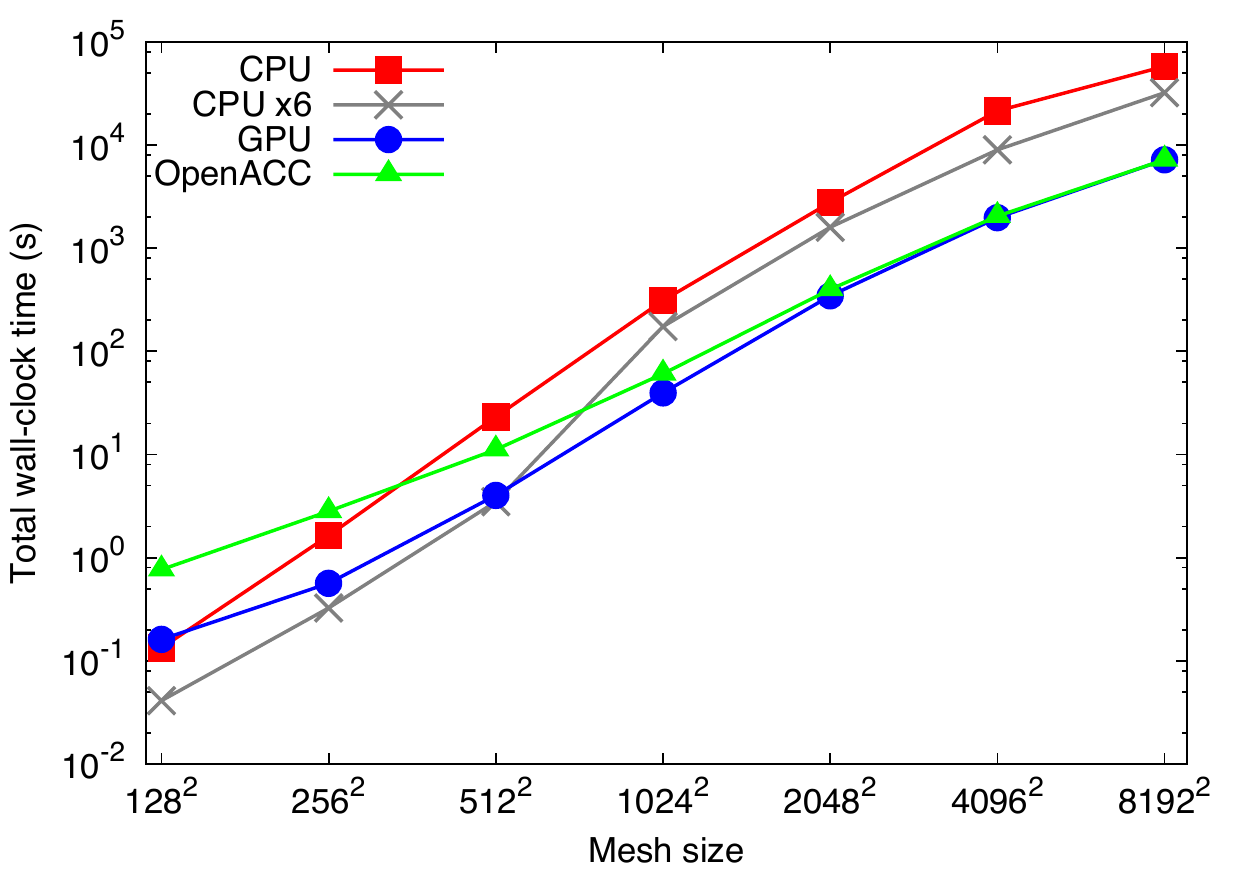}
\caption{Performance comparison of the CPU Laplace solver on one core and six cores (via OpenMP), the fully optimized CUDA-based GPU solver, and OpenACC-based GPU solver, for a wide range of mesh sizes.}
\label{fig:laplace_cpu-gpu}
\end{center}
\end{figure}

Next, we compared the performance of the single-core CPU, six-core CPU, fully optimized CUDA-based GPU, and OpenACC-based GPU solvers over a wide range of mesh sizes for double-precision calculations, shown in Fig.~\ref{fig:laplace_cpu-gpu}. At mesh sizes of 1024\textsuperscript{2} and above, the GPU solver ran faster than the CPU solver on either one or six cores; at most, the GPU solver performed up to about 10 and 4.6 times faster than the single-core and six-core CPU solvers, respectively, for double precision. The single-precision code performed similarly: the CUDA-based GPU solver ran about 16 and 4.6 faster than the single- and six-core CPU versions, at best.

At smaller mesh sizes, the OpenACC implementation ran nearly five times slower than the native, fully optimized GPU version, but as the mesh size increased this gap decreased to nearly zero, demonstrating almost equal performance at the largest grid sizes. This behavior was replicated in single-precision calculations, although the OpenACC solver performed about 8\% slower at the largest grid sizes and 6.9 times slower at the smallest mesh sizes.

Note that we did not optimize the block size for the GPU solver, but left it constant for this simple demonstration. Similarly, we allowed the OpenACC compiler to determine the optimal configuration, rather than manually adjust its equivalents (``gang'' and ``vector'' sizes). More in-depth optimization could further improve performance for both implementations.

Finally, we observed that while the native fully optimized GPU solver showed the best performance for larger grid sizes, the OpenACC version performed nearly as well, especially for double-precision calculations. These results suggest that OpenACC is a good alternative to writing native GPU code, especially considering the fact that porting CPU applications to CUDA can require days to weeks of work while adding the OpenACC compiler directives takes only hours or even minutes. However, OpenACC support is currently limited, and can only be applied to applications that already favor parallelization---such as those based on loops with independent iterations---and where functions may be inlined. With this in mind, OpenACC is a good choice to quickly accelerate existing code and determine potential speedup, while writing native GPU applications offers the highest potential performance if fully optimized. In either case, algorithms may need to be redesigned in order to support massive parallelization, although this was not necessary in the current example.

\subsection{Lid-driven cavity flow}
\label{sec:cavity}

\subsubsection{Methodology}

The second case study consisted of solving the two-dimensional, laminar incompressible Navier--Stokes equations based on the finite difference method, using the solution procedure given by Griebel et al.~\cite{Griebel:1998}. The domain was discretized with a uniform, staggered grid (i.e., the pressure values are located at the centers of grid cells while velocity values are located along the edges). Briefly, the discretized momentum (assuming no gravity\slash body-force terms) and pressure-Poisson equations are:
{\allowdisplaybreaks \begin{IEEEeqnarray}{rCl}
	F_{i,j}^{(n)} & = & u_{i,j}^{(n)} \label{eq:Fij} \\ 
	& & +\: \delta t \left( \frac{1}{\text{Re}} \left( \left[ \ddydxx{u}{x} \right]_{i,j}^{(n)} + \left[ \ddydxx{u}{y} \right]_{i,j}^{(n)} \right) - \left[ \dydx{(u^2)}{x} \right]_{i,j}^{(n)} - \left[ \dydx{(uv)}{y} \right]_{i,j}^{(n)} \right) , \IEEEeqnarraynumspace \nonumber \\ 
	\IEEEeqnarraymulticol{3}{r}{i = 1, \ldots, i_{\max} - 1, \quad j = 1, \ldots, j_{\max}} , \nonumber \\
	G_{i,j}^{(n)} & = & v_{i,j}^{(n)} \label{eq:Gij} \\
	& & +\: \delta t \left( \frac{1}{\text{Re}} \left( \left[ \ddydxx{v}{x} \right]_{i,j}^{(n)} + \left[ \ddydxx{v}{y} \right]_{i,j}^{(n)} \right) - \left[ \dydx{(uv)}{x} \right]_{i,j}^{(n)} - \left[ \dydx{(v^2)}{y} \right]_{i,j}^{(n)} \right) \nonumber , \\
	\IEEEeqnarraymulticol{3}{r}{i = 1, \ldots, i_{\max}, \quad j = 1, \ldots, j_{\max} - 1} , \IEEEeqnarraynumspace \nonumber
\end{IEEEeqnarray}}%
{\allowdisplaybreaks \begin{IEEEeqnarray}{rCl}
	\frac{p_{i+1,j}^{(n+1)} - 2 p_{i,j}^{(n+1)} + p_{i-1,j}^{(n+1)}}{(\delta x)^2} & + & \frac{p_{i,j+1}^{(n+1)} - 2 p_{i,j}^{(n+1)} + p_{i,j-1}^{(n+1)}}{(\delta y)^2} \label{eq:pressure} \\
	& & = \frac{1}{\delta t} \left( \frac{F_{i,j}^{(n)} - F_{i-1,j}^{(n)}}{\delta x} + \frac{G_{i,j}^{(n)} - G_{i,j-1}^{(n)}}{\delta y} \right) \IEEEeqnarraynumspace \nonumber \\
	\IEEEeqnarraymulticol{3}{r}{i = 1, \ldots, i_{\max}, \quad j = 1, \ldots, j_{\max}} , \IEEEeqnarraynumspace \nonumber \\
	\IEEEeqnarraymulticol{3}{C}{u_{i,j}^{(n+1)} = F_{i,j}^{(n+1)} - \frac{\delta t}{\delta x} \left( p_{i+1,j}^{(n+1)} - p_{i,j}^{(n+1)} \right) , } \label{eq:uij} \\
	\IEEEeqnarraymulticol{3}{r}{ i = 1, \ldots, i_{\max} - 1, \quad j = 1, \ldots, j_{\max}}, \IEEEeqnarraynumspace \nonumber \\
	\IEEEeqnarraymulticol{3}{C}{v_{i,j}^{(n+1)} = G_{i,j}^{(n+1)} - \frac{\delta t}{\delta y} \left( p_{i,j+1}^{(n+1)} - p_{i,j}^{(n+1)} \right) , } \label{eq:vij} \\
	\IEEEeqnarraymulticol{3}{r}{i = 1, \ldots, i_{\max}, \quad j = 1, \ldots, j_{\max} - 1}, \IEEEeqnarraynumspace \nonumber%
\end{IEEEeqnarray}}%
where \emph{i} and \emph{j} indicate the \emph{x} and \emph{y} cell coordinates, respectively, $(n)$ indicates the time step (i.e., corresponding to time $t_n$), \emph{u} and \emph{u} are the velocity components in the $x$ and $y$ directions, respectively, $p$ is the pressure, Re is the Reynolds number, and $\delta t$ is the time-step size. The second derivatives in Eqs.~\eqref{eq:Fij} and \eqref{eq:Gij} were treated with central differences, while the first derivatives were treated with a mixture of central differences and the donor-cell discretization (see~Griebel et al.~\cite{Griebel:1998} for details).

In this example, we used the above procedure to solve the case of lid-driven flow in a square cavity, corresponding to no-slip boundary conditions on the vertical sides and bottom and a unit horizontal velocity along the top. Boundary conditions were treated numerically as described by Griebel et al.~\cite{Griebel:1998}. As with the Laplace solver above, we used boundary cells to avoid conditional statements and the associated potential thread divergence. Kernel functions evaluated the boundary conditions for both velocity and pressure values.

The GPU code contained 11 kernel functions: one to set the velocity boundary conditions, two corresponding to Eqs.~\eqref{eq:Fij} and \eqref{eq:Gij}, one to calculate the $L^2$-norm of the pressure (for a relative SOR tolerance), two to set the horizontal and vertical pressure boundary conditions, two for the red and black portions of the SOR algorithm solving Eq.~\eqref{eq:pressure}, one to calculate the pressure residual for each SOR iteration, and two corresponding to Eqs.~\eqref{eq:uij} and \eqref{eq:vij} (which also return the maximum \emph{u}- and \emph{v}-velocities). All memory remained on the GPU during the simulation (e.g., arrays holding $F_{ij}$, $p_{\text{black}}$, $u_{ij}$), except for that needed to evaluate the stopping criterion for the SOR iteration and the maximum velocities to calculate the time-step size based on stability criteria (see Griebel et al.~\cite{Griebel:1998}). Learning from our experiences with the Laplace solver, we used shared memory for these global reduction operations such that only one value per block needed to be transferred back to the CPU. Performance timing included all these memory transfers, including those needed to initialize all variables on the GPU at the beginning and return the pressure and velocity values at the end of the simulation.

In the same manner as the Laplace solver, the OpenACC solver was based on the CPU solver, with directives instructing the compiler to use the GPU on loops matching the kernel functions of the GPU solver. The OpenMP solver was created the same way, using the appropriate compiler directives. No specific optimization instructions were given to either the OpenMP or OpenACC solvers; rather, we allowed the compiler to manage this automatically.

In order to study the performance of the GPU and OpenACC solvers against the CPU versions, we varied the mesh size from 64\textsuperscript{2} to 2048\textsuperscript{2}. The source code was written in standard and CUDA C for the CPU and GPU versions, respectively, with compiler directives added to the CPU version to create OpenMP and OpenACC versions.\footnote{The full source code is available online: \href{http://github.com/kyleniemeyer/lid-driven-cavity_gpu}{\nolinkurl{http://github.com/kyleniemeyer/lid-driven-cavity_gpu}}} In the native, CUDA-based GPU version, we kept a constant block size of 128 $\times$ 1 except for the cases of the mesh size being 64\textsuperscript{2} or 128\textsuperscript{2}, where we used half the mesh size in one direction (e.g., 32 $\times$ 1 for a mesh size of 64\textsuperscript{2}). Based on our experience with the Laplace solver, we used grid configurations of $\nicefrac{N}{2 B} \times N$ for the pressure solver and $\nicefrac{N}{B} \times N$ for the velocity portions, where \emph{N} is the number of mesh cells in one direction (e.g., 512) and \emph{B} is the block size (e.g., 128).

For each mesh size, we performed a single time step integration---the initial time step. Due to the stability criteria for selecting the time-step size, the finer mesh sizes required increasingly smaller time-step sizes. All calculations were performed in double precision.

\subsubsection{Results}

\begin{figure}[tbp]
\begin{center}
\includegraphics[width=.8\linewidth]{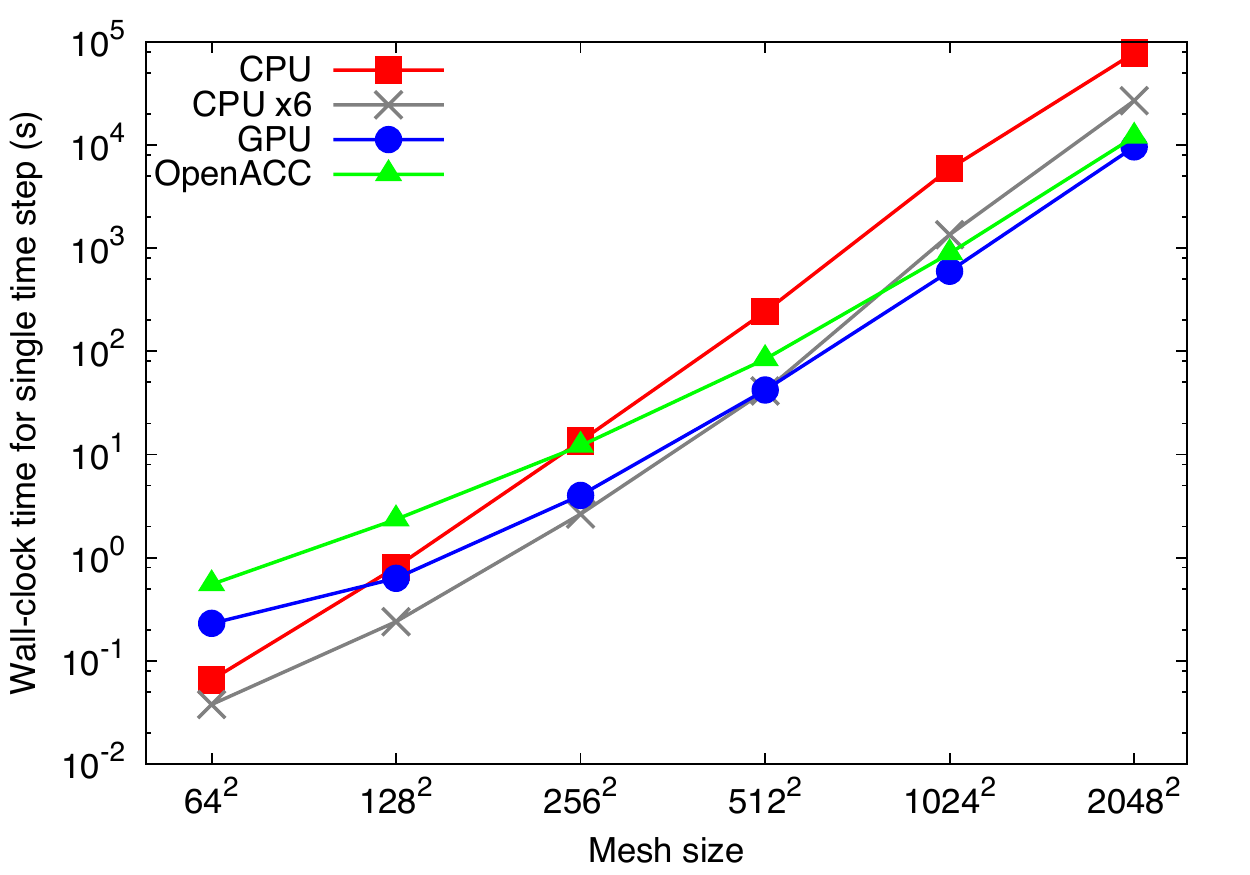}
\caption{Performance comparison of a single time step of the lid-driven cavity problem using the CPU solver on one core and six cores (via OpenMP), the fully optimized CUDA-based GPU solver, and OpenACC-based GPU solver, for a wide range of mesh sizes.}
\label{fig:cavity_cpu-gpu}
\end{center}
\end{figure}

Figure~\ref{fig:cavity_cpu-gpu} shows the performance comparison of the solvers over a wide range of mesh sizes, for a single time step. Qualitatively, the results demonstrated similar trends to those of the Laplace solver shown in Fig.~\ref{fig:laplace_cpu-gpu}. This was due to the computational intensity of the red-black SOR solution of the pressure-Poisson equation, which was solved in the same manner in both the Laplace solver and this finite-difference-based Navier--Stokes solver. At mesh sizes smaller than 128\textsuperscript{2}, the single- and six-core CPU solvers ran faster than either GPU solver, but with increasing mesh size the native, optimized GPU solver became 8.1 and 2.8 times faster than the single- and six-core CPU solvers, respectively, at a mesh size of 2048\textsuperscript{2}.

The OpenACC implementation for this problem also behaved similarly to that in the Laplace case study. With increasing problem size, the OpenACC solver approached the performance of the native GPU version, running 1.3 times slower than the native GPU code at the largest problem size. At smaller problem sizes, it ran nearly four times slower than the native GPU version. It is clear, however, that OpenACC offers nearly the same performance as native GPU code at large problem sizes.

\section{GPUs in computational fluid dynamics}

The heavy computational demands of high-fidelity fluids simulations typically prevent industrial or academic researchers from performing and using such studies. CFD applications in particular stand to benefit from GPU acceleration due to the inherent data parallelism of calculations for both finite difference and finite volume methods, and as GPU hardware and software matured more researchers exploited the application of GPU computing in their respective areas of interest. While Vanka et al.~\cite{Vanka:2011vc} reviewed some of the literature on using GPUs for CFD applications, we attempt to provide a more comprehensive survey and discuss recent advances, particularly for laminar-, turbulent-, and reactive-flow modeling. In addition, we review efforts to accelerate solvers based on the lattice Boltzmann method for simulating fluid flows.

\subsection{Laminar flow}
\label{sec:laminar}

We first survey efforts to develop GPU-accelerated laminar flow solvers, both incompressible and compressible. Most of the early work involved transferring some or all calculations to the GPU, leaving the CPU to initialize and drive the simulation. This can minimize expensive memory transfer between the CPU and GPU, but balancing the loads on each processor is important for optimal performance. In other words, optimal codes will avoid leaving either processor idle. In addition, most GPU-accelerated CFD codes were limited to structured meshes, with notable recent exceptions to be discussed later.

Bolz et al.~\cite{Bolz:2003cz}, Harris and coworkers~\cite{Harris:2003vh,Harris:2004wc}, and Kr\"{u}ger and Westermann~\cite{Kruger:2003bg} first implemented real-time, physics-based fluids simulation on GPUs using the stable fluids approach of Stam~\cite{Stam:1999ey} for solving the incompressible Navier--Stokes equations. Liu et al.~\cite{Liu:2004vl} used the same approach to solve flow around complex boundaries. In these approaches, the finite difference equations were parallelized such that the discretized derivatives for each grid point were solved concurrently. In other words, instead of looping over each point in a serial fashion, the derivatives were solved simultaneously. However, the stable fluids approach includes excessive numerical dissipation---it is limited to physics-based animations rather than accurate, numerical simulations.

Following these initial efforts, researchers began to develop more accurate Navier--Stokes solvers for GPUs. Scheidegger et al.~\cite{Scheidegger:2005we} demonstrated a GPU-based incompressible flow solver using the Simplified Marker and Cell approach on a structured grid. This work was performed before the release of CUDA, so Scheidegger et al.~\cite{Scheidegger:2005we} used the graphics programming languages OpenGL and Cg, and stored data structures (e.g., velocities, pressure, temperature) in GPU texture memory elements, such as pixel buffers, which are used in GPUs to perform graphics rendering tasks. On the classical lid-driven cavity problem, the GPU performed up to 21 times faster than the CPU at a Reynolds number of 1000 and a grid size of 128\textsuperscript{2}. They also demonstrated their code using simulations of flow through a domain with obstacles, flow through a wind tunnel with vehicle object, smoke trails, and natural convection with heated walls. Interestingly, the GPU simulations ran fast enough to allow real-time flow visualization, which was made easier since the graphics pipeline was already being utilized for flow calculations.

Hagen et al.~\cite{Hagen:2006uq} were the first to develop a GPU solver for the three-dimensional Euler equations using a high-resolution finite volume method. The GPU executed the cell flux evaluation and time integration, implemented using the Cg and OpenGL programming languages, while the CPU drove the calculation and evaluated the time-step size based on stability considerations. On the GPU, grid cells and their associated properties were represented as fragments (pixels that have not been displayed onscreen), while combinations of grid cell properties (e.g., cell averages, fluxes) were given as textures, processed using fragment shaders (i.e., kernels). Hagen et al.~\cite{Hagen:2006uq} compared their GPU solver against a mature CPU solver using a number of test cases, including a two-dimensional bubble-shock interaction and three-dimensional Rayleigh--Taylor instability, and showed a speedup of more than 10 times for both cases. For their tests, the CPU code was highly optimized while the GPU code was not, suggesting that greater acceleration might be possible.

Brandvik and Pullan~\cite{Brandvik:2007bn,Brandvik:2008up} also developed two- and three-dimensional GPU solvers for the compressible, inviscid Euler equations, which they used to simulate flows through turbines. This was one of the first CFD applications to use the general-purpose programming languages BrookGPU and CUDA for the two- and three-dimensional solvers, respectively, rather than the specialized graphics languages of earlier studies. Using the finite volume approach, the controlling CPU handled pre- and post-processing while the GPU performed the actual computations. For example, the CPU constructed the grid and evaluated face areas and cell volumes, while the time steps (e.g., evaluating cell fluxes) were performed on the GPU. In the BrookGPU implementation, texture memory contained node information and kernels performed the computations on these values (e.g., flux calculation). With CUDA, the global grid was split into smaller three-dimensional chunks associated with thread blocks, using the efficient shared memory to store cell values. In this case, the shared memory of each block could hold \SI{16}{\kilo\byte}, so a typical sub-grid size was 16 $\times$ 10 $\times$ 5. The BrookGPU-based two-dimensional solver was evaluated using a study of transonic flow over a turbine, showing a speedup of 29 times over an equivalent Fortran CPU code; both the CPU and GPU codes produced identical results. The CUDA-based three-dimensional solver was tested on simulation of flow over a low-speed linear turbine cascade and showed a 16 times speedup over an equivalent single-core CPU version.

Elsen et al.~\cite{Elsen:2008jv} transferrred the steady, compressible Euler portions of the Navier--Stokes Stanford University Solver to the GPU, using BrookGPU. Elsen et al.~\cite{Elsen:2008jv} first tested their GPU-accelerated code on a simulation of flow over the traditional NACA 0012 airfoil, showing a 14 times speedup for third-order accuracy compared with optimized code running on a single CPU core. They also demonstrated the GPU code on a study of hypersonic flow (Mach 5) over a vehicle, considering cases with \num{720000} and 1.5 million nodes. In this case, the GPU solver ran about 16 times faster than the CPU code, reducing the wall-clock time from around four hours for CPU to only 15 minutes for GPU.

Molemaker et al.~\cite{Molemaker:2008vy}, hoping to improve upon the stable fluids approach used in animation, developed a solver for the simulation of incompressible flows using the quadratic upwind interpolation for convection kinetics (QUICK) advection scheme and a projection method to solve the Poisson pressure equation. QUICK was chosen to reduce numerical dissipation. Cohen and Molemaker~\cite{Cohen:2009tp} later showed a second-order, double precision, finite volume code solving the incompressible flow equations with the Boussinesq approximation. Both codes ran nearly an order of magnitude faster than equivalent CPU versions, noting that Cohen and Molemaker~\cite{Cohen:2009tp} performed their comparison in parallel on an eight-core CPU.

Phillips et al.~\cite{Phillips:2009vj} also developed a GPU-based solver for the compressible Euler equations, and additionally were the first to accelerate a portion of an existing CPU-based multi-block CFD solver, MBFLO, by moving the unsteady flow and laminar stress calculations to the GPU. The Euler code, simulating subsonic flow through a nozzle, ran up to 20 times faster on a single GPU against an equivalent version running on a single CPU. The GPU-accelerated MBFLO, simulating low-speed flow over a cylinder, also showed nearly the same improvement, 14 times, compared against the CPU-only version. Phillips et al.~\cite{Phillips:2009vj} found that (1) for the Euler solver, the GPU code became more efficient (i.e., able to process more grid cells in the same time) as the domain size was increased, while the CPU became less efficient; and (2) the performance enhancement of the GPU-accelerated MBFLO improved as more subroutines were moved to the GPU from the CPU. The boundary condition and block-to-block communication routines were the only portions not accelerated.

Shinn and Vanka~\cite{Shinn:2009ui} implemented a multigrid method using the semi-implicit method for pressure linked equations (SIMPLE) algorithm to solve the incompressible Navier--Stokes equations. Using the classical lid-driven cavity problem, their GPU code performed over an order of magnitude faster than the CPU version for finer meshes (e.g., 512\textsuperscript{2}). Steady-state calculations with an extremely fine mesh (4096\textsuperscript{2}) could be performed on the GPU in around one minute.

Thibault and Senocak~\cite{Thibault:2009uq} developed a CUDA-based GPU solver for the incompressible Navier--Stokes equations. Similar to the work of Phillips et al.~\cite{Phillips:2009vj}, this code could run on multiple GPUs by decomposing the domain into smaller blocks of cells. Following the trend of validation using the lid-driven cavity problem, they demonstrated a 13 times speedup running their code on one GPU compared against an equivalent version running on one CPU core. They also compared the performance of the accelerated code on multiple GPUs against the serial version running on a single CPU core; however, this is an unfair comparison, equivalent to comparing performance on a CPU cluster against a single CPU. The same group extended their code to run on multiple clusters of GPUs using Message Passing Interface (MPI). In this case, the performance of multiple GPUs was compared against parallelized code running on an eight-core CPU. A single GPU ran 11 times faster than the eight-core CPU, while eight GPUs performed up to 68 times faster.

Griebel and Zaspel~\cite{Griebel:2010kb} developed the first GPU solver capable of simulating two-phase unsteady three-dimensional incompressible flows, based on a level-set approach. This code was based on an existing in-house CPU solver, NaSt3DGPF, which discretized the Navier--Stokes equations using a finite-different approach. The most computationally expensive parts of the CPU code were ported to the GPU: (1) the Jacobi-preconditioned conjugate gradient solver for the Poisson equation and (2) the reinitialization process for the level set function. Simulating an air bubble rising through water, the GPU code performed about 10 times faster than the original CPU version, comparing one GPU to one CPU core with a grid resolution of 300\textsuperscript{3}.

Recently, Zaspel and Griebel~\cite{Zaspel:2012gp} updated their solver to run fully on multiple GPUs---rather than just the Poisson solver and level-set reinitialization portions. This eliminated most CPU-GPU memory transfer during the simulation and led to a more than 30\% improvement in performance over their previous GPU-accelerated code~\cite{Griebel:2010kb}. In addition, they used MPI to parallelize over multiple GPUs, overlapping most communication with computations to hide the data transfer time. Zaspel and Griebel~\cite{Zaspel:2012gp} advocated not only measuring the raw performance of code on one GPU against one CPU socket (i.e. multiple cores), rather than the common comparison of one GPU vs.\ a single CPU core, but also comparing performance normalized by hardware cost and power consumption (performance per dollar and watt, respectively). Using the same benchmarking problem as in their previous study, Zaspel and Griebel~\cite{Zaspel:2012gp} found that their new code performed around three times faster on a single NVIDIA Tesla c2050 GPU than on two six-core CPUs, together priced similarly to the GPU. In addition, the single GPU consumed less than half the power of both CPUs combined---with a total of 12 cores. They also reported the scalability of the GPU code on a cluster of many GPUs, showing nearly 20 times the performance on 48 GPUs than on a single GPU for an overall problem size of 256\textsuperscript{3}. The somewhat low scaling efficiency of around 40\% might be explained by the small problem size each GPU handled (around 70\textsuperscript{3}); GPUs tend to perform better compared to CPUs on larger problem sizes.

Up to this point, all of the GPU-accelerated CFD codes discussed here relied on structured grids, limiting the complexity of potential objects in the domain. Corrigan et al.~\cite{Corrigan:2011ig} developed the first GPU-accelerated solver for unstructured grids, capable of simulating three-dimensional compressible, inviscid flows. Structured grids allow easier optimization due to regular memory access patterns; Corrigan et al.~\cite{Corrigan:2011ig} used the memory coalescing feature of modern GPUs, where high memory access performance can be achieved by ensuring that consecutive threads access consecutive memory locations. They realized this by ensuring that neighboring elements were located consecutively in memory, using a bin numbering scheme. In addition, in order to limit potential thread divergence by the necessarily different treatment of boundary elements compared to neighboring non-boundary elements, boundary elements were stored consecutively in memory. Corrigan et al.~\cite{Corrigan:2011ig} first tested their code on a simulation of supersonic flow over a NACA 0012 airfoil, showing an average performance speedup of about 32 and 9 compared against a CPU code running on one core and four cores, respectively. In addition, they simulated supersonic flow over a missile, demonstrating similar speedups of about 34 and 10, respectively. In both cases, the number of elements ranged from around \num{100000} to nearly two million.

Block et al.~\cite{Block:2012gf} developed a GPU-accelerated compressible flow solver using the finite volume evolution Galerkin method with both regular and adaptive meshes, targeted at geophysical flow simulations. Rather than the entire code, they transferred only the most expensive portion---the evolution Galerkin operator---to the GPU. Compared to a single CPU core, this procedure ran nearly 30 times faster on the GPU, resulting in an overall code speedup factor of around six for simulations with over \num{16000} cells. Block et al.~\cite{Block:2012gf} demonstrated their code using simulations of free convection: (1) a warm air bubble and (2) a small cold air bubble on top of a larger warm bubble.

Following the work of Hagen et al.~\cite{Hagen:2006uq}, Brandvik and Pullan~\cite{Brandvik:2007bn,Brandvik:2008up}, and Phillips et al.~\cite{Phillips:2009vj}, Lefebvre et al.~\cite{Lefebvre:2012} developed GPU solvers for the two-dimensional and three-dimensional Euler equations on structured grids. In order to achieve higher performance, they undertook a number of specific implementation techniques, including (1) favoring redundant computation over global memory access, which improved performance by about 20--30\%; (2) optimizing thread block dimensions to maximize thread occupancy; and (3) the use of pinned memory to reduce CPU-GPU memory transfer times by a factor of three. Lefebvre et al.~\cite{Lefebvre:2012} demonstrated their two-dimensional solver using simulations of a shock tube and supersonic wind tunnel, showing speedup factors of 60---for the second-order spatial discretization---against a single CPU core. Their three-dimensional Euler solver ran 10 times faster on a single GPU than an equivalent version on a six-core CPU. All cases were performed using single precision computations.

In addition to the traditional Navier--Stokes solver based on the finite difference or finite volume methods, some groups used alternative approaches. Ran et al.~\cite{Ran:2011} developed a GPU solver for the one-dimensional Euler equations using the conservation element and solution element (CESE)~\cite{Chang:1995} method. CESE is a different approach to solving conservation laws that is especially useful for flow problems with discontinuities (e.g., shocks, boundaries), where it doesn't require as fine a grid as traditional approaches such as the finite difference method. Ran et al.~\cite{Ran:2011} used shared memory to store data from the previous time step---needed to perform the time-step integration---but this approach posed difficulties as points on the boundaries of blocks need to access this data in the adjacent block, while shared memory is private to each block. To solve this, they organized the thread blocks to contain an extra thread for use as a cache, storing in the block's shared memory the data needed at the inter-block boundary location. Simulating condensation in a shock tube, the GPU-based CESE solver performed up to 71 times faster than a single CPU core executing a serial version.

Arguing that traditional solution methods do not work well on the thread parallelism of GPUs, Kuo et al.~\cite{Kuo:2011jo} developed a new approach: the split Harten, Lax and van Leer (SHLL) method. The original HLL method~\cite{Harten:1983} computes fluxes across cell interfaces by assuming the presence of two propagating waves and integrating their governing equations. The SHLL variant modifies the flux expressions into a vector-split form that is ideal for parallel computation---particularly on GPUs---due to higher locality. In order to achieve better performance, Kuo et al.~\cite{Kuo:2011jo} implemented their solver such that after initialization the entire simulation ran on the GPU without any interaction---and therefore no memory transfer---with the CPU. However, this approach might not be feasible in more traditional CFD solution methods, where global reduction operations may be necessary (e.g., to stop Jacobi\slash GS iterations, or determine the next time-step size based on fluid velocity values). Demonstrated using the (1) Euler equations for a simulation of two-dimensional shock/bubble interaction and (2) inviscid shallow equations for a simulation of a two-dimensional dam break problem, the GPU solver performed more than 60 times faster for larger meshes, compared with a single CPU core.

Chabalko et al.~\cite{Chabalko:2013bo} implemented a GPU-based solver for the two-dimensional unsteady vortex lattice method, used to compute the inviscid aerodynamic forces and resulting flow fields around solid bodies (e.g., airfoils, aircraft, wind turbines). This method is based on the conservation of circulation, and focuses on tracking the circulation of bound and free vortices. Chabalko et al.~\cite{Chabalko:2013bo} used the GPU to accelerate the most computationally intensive portion of the method: the evaluation of the influence of all free vortices on all other free vortices (governed by the Biot--Savart law). Using simulations of (1) a flat plate moving in a cnoidal ground effect and (2) the roll up of a vortex filament, they demonstrated nearly two orders of magnitude speedup over an equivalent serial CPU version.

\subsection{Turbulent flow}
\label{sec:turbulent}

Most real-world fluid flow problems, and in particular those relevant to engineering applications such as flow through turbomachinery and engines, and over cars and aircraft, are turbulent. However, due to the difficulty in transferring CFD codes to operation on graphics processors, only recently have researchers begun to accelerate turbulent-flow simulations using GPUs. Ironically, this is one of the areas that need acceleration the most due to the added expense of accurate turbulence models. Furthermore, reactive flow simulations are arguably in greater need of speedup---this will be discussed in the next section.

Phillips et al.~\cite{Phillips:2010tn} developed one of the first GPU solvers capable of simulating turbulence using the $k$--$\omega$ model, extending on the group's previous work porting portions of the existing MBFLO solver to the GPU~\cite{Phillips:2009vj}. In addition, their new solver was capable of running on a cluster of multiple CPU\slash GPU nodes, using a domain decomposition technique to give each node responsibility for a block of the overall domain. The entire time-step loop, including the calculations of laminar stress, turbulent viscosity coefficient, and cell flux integration, was performed on the GPU in order to minimize slow CPU-GPU memory transfer. The CPU only drove the simulation and passed information between the blocks of the domain, using MPI to transfer information between independent cluster nodes. Phillips et al.~\cite{Phillips:2010tn} also improved performance by implementing a novel asynchronous memory transfer using CUDA streams; in their previous work, the GPU remained idle while the CPU transferred memory between different blocks (i.e., subdomains). Here, each block was further divided in half such that the GPU could continue to perform calculations on one half while the CPU transferred memory associated with the other half of the block; this improved performance up to 40\%. Phillips et al.~\cite{Phillips:2010tn} tested their code using a simulation of unsteady turbulent flow over a cylinder, finding that a cluster of eight GPUs performed about nine times faster than an equivalent parallel code running on eight quad-core CPUs.

Jespersen~\cite{Jespersen:2010} accelerated the existing turbulent CFD code OVERFLOW by moving a portion of the code to the GPU. OVERFLOW solves the Reynolds-averaged Navier--Stokes equations using the finite difference method, with an implicit time-stepping scheme to avoid stability issues. Jespersen~\cite{Jespersen:2010} ported the Jacobian algorithm used to solve the resulting large, sparse linear system to the GPU, while the rest of the code remained on the host CPU. This hybrid solver demonstrated a 25\% improvement in performance by using the GPU and CPU together, over the CPU alone.

Kampolis et al.~\cite{Kampolis:2010jf} developed a GPU-accelerated unstructured grid-based turbulent flow solver, modeling turbulence using the one-equation Spalart--Allmaras model, and integrated it within an evolutionary algorithm-based optimization algorithm. This approach differed from the unstructured grid work of Corrigan et al.~\cite{Corrigan:2011ig} in that this implementation is based on the vertex-centered finite difference approach. They also implemented a mixed-precision version of their solver that offered nearly the same level of accuracy as double- but with the speed approaching that of single-precision. The GPU-accelerated CFD code performed up to around 28, 25.2, and 19.6 times faster than a serial CPU code for single, mixed, and double precision, respectively, for two-dimensional flow over a NACA 4415 airfoil. Simulations of three-dimensional flow around an aircraft and within a supersonic compressor cascade performed similarly. This efficient GPU code was then used to optimize airfoil shapes, showing the potential utility of GPU-accelerated applications to design. Asouti et al.~\cite{Asouti:2010gy} extended this unstructured grid approach, restructuring the code and investigating schemes to optimize memory access on the GPU. They reported a maximum factor of speedup of approximately 45 compared to equivalent Fortran code running on a single CPU core, for double-precision calculations.

Shinn et al.~\cite{Shinn:2010wf} developed the first (to our knowledge) direct numerical simulation (DNS) solver for the GPU, using it to study turbulent flow through a square duct. The solution of the pressure-Poisson equation was performed on the GPU, using the red-black GS scheme (a parallel version of the traditional GS, as discussed in Section~\ref{sec:laplace_meth}). A multigrid algorithm enhanced the convergence of the Poisson equation solution; each iteration was performed on the GPU. In order to reduce global memory access, they used texture memory to fetch the pressure values during the Poisson solution, improving performance by 10\%. The GPU-accelerated solver performed over an order of magnitude faster than the equivalent CPU version, with increasing speedup as the mesh size increased (from about 12 times using a 128 $\times$ 32\textsuperscript{2} mesh to nearly 16 times with a 512 $\times$ 128\textsuperscript{2} mesh).

Following this effort, Shinn and Vanka~\cite{Shinn:2011vq} also demonstrated the first GPU solver capable of performing large-eddy simulation (LES) of turbulent incompressible flows. Unfortunately, they did not provide details on their GPU implementation, or compare the performance against an equivalent CPU version. However, their approach appears to be based their earlier DNS work~\cite{Shinn:2010wf} described above, using the same discretization and integration schemes and the red-black GS method for the pressure-Poisson equation.

Alfonsi et al.~\cite{Alfonsi:2011kn} also developed a GPU-based DNS solver for incompressible flow. Their discretization was based on a mixed spectral-finite difference approach. The GPU performed the core of the algorithm, including the fast Fourier transform used in the spectral portion of the approach to transform the velocity field to and from spectral space. They optimized memory access by (1) ensuring coalesced access throughout, which involved modifying the Thomas (i.e., tridiagonal matrix) algorithm for the Poisson equation; and (2) using constant memory to store the diagonal matrices used in that algorithm. Alfonsi et al.~\cite{Alfonsi:2011kn} used their code to study wall-bounded turbulence in a plane-channel domain, with a mesh of size 256\textsuperscript{2} $\times$ 181. The GPU-accelerated solver performed nearly 25, 13.2, and 7.8 times faster than single-, two-, and four-CPU versions, respectively.

Brandvik and Pullan~\cite{Brandvik:2011dt}, building upon their earlier work described above, designed a new three-dimensional Navier--Stokes solver for graphics processors, TURBOSTREAM, based on an existing class of CPU codes for simulating flows in turbomachinery. In this solver, the equations were discretized using a finite-volume method with a structured grid; turbulence was modeled with a simple mixing-length model. The new code---executed entirely on the GPU---ran approximately 19 times faster on a single GPU than the original code on a quad-core CPU, using a grid with one million nodes. The code was also validated with a high-resolution simulation (4.5 million grid cells) of steady-state flow through a three-stage turbine with leakage paths, executing in less than 10 minutes on a cluster of four GPUs.

DeLeon and coworkers~\cite{DeLeon:2012wu,DeLeon:2013vp} recently demonstrated another GPU-based solver for LES of turbulent incompressible flows. The subgrid-scale terms were modeled using the Lagrangian dynamic Smagorinsky model. DeLeon et al.~\cite{DeLeon:2013vp} parallelized their solver to run on a cluster of multiple GPUs using MPI, such that the overall code contained two levels of parallelism. The overhead of communication between blocks was minimized by using the same asynchronous memory transfer as Phillips et al.~\cite{Phillips:2010tn}. They did not compare the performance of their LES GPU solver to an equivalent CPU version, but simulations of turbulent channel flow with approximately 9.4 million grid cells took 45 hours to complete running on a cluster with eight total GPUs. However, previous comparisons of their flows solver with the lid-driven cavity benchmarking problem showed that the GPU version running on two Tesla S1070 cards ran 26 times faster than the version running in parallel on eight CPUs~\cite{Jacobsen:2010vt,Thibault:2012kg}. Jacobsen and Senocak~\cite{Jacobsen:2013ik} discussed in detail the multi-level parallelization strategies used in these efforts for incompressible flow simulations, including hybrid MPI-CUDA and MPI-OpenMP-CUDA.

Iman Gohari et al.~\cite{Gohari:2012bd} developed a solver for the incompressible, turbulent Navier--Stokes equations, where both the grid generation stage and flow solver were performed on the GPU, and used it to simulate flow over airfoils. Using the stream function-vorticity formulation, they modeled turbulence using the Balwin--Lomax closure method. Systems of linear equations present in both stages were solved using Jacobi iteration, which, unlike the GS method, does not require special treatment for parallelization, although convergence may occur more slowly. As with other approaches using structured grids, Iman Gohari et al.~\cite{Gohari:2012bd} exploited global memory coalescing to improve performance, and focused on optimizing memory performance in detail. Using single-precision calculations, they demonstrated an overall speedup (grid construction + flow solution) of up to around 40, comparing a single GPU and single CPU core.

Recently, a number of groups developed GPU-based DNS solvers. Salvadore et al.~\cite{Salvadore:2012} demonstrated such a code for turbulent, compressible flows, based on the finite-difference method. Profiling of the original serial CPU version revealed that the code spent more than 90\% of the computational time on calculating the convective and viscous fluxes---however, to minimize CPU-GPU memory transfer, rather than transferring just the most expensive portions, all of the code was moved to the GPU. In order to improve the performance of these sections on the GPU, Salvadore et al.~\cite{Salvadore:2012} presented a number of optimization steps, including coalescing global memory through shared memory use, eliminating use of slower local memory, and increasing the amount of work per thread. These optimization steps increased the speedup factor over a quad-core CPU from 4.5 to 22. They used the final GPU solver to perform DNS simulations of a spatially evolving compressible mixing layer, with a grid consisting of 1536 $\times$ 128 $\times$ 140 points; a total of eight quad-core CPUs were necessary to meet the performance of a single GPU, while consuming nearly 2.4 times the power.%

Khajeh-Saeed and Perot~\cite{KhajehSaeed:2013kz} also recently developed a DNS solver for GPU clusters. The Poisson pressure equation was solved using a polynomial-preconditioned conjugate gradient (CG) method, with GPU-specific optimizations such as using shared memory to reduce non-coalesced global memory access during the Laplacian matrix evaluation. They used MPI for inter-GPU communication, overlapping GPU computation with CPU-GPU and MPI communications. Khajeh-Saeed and Perot~\cite{KhajehSaeed:2013kz} compared the performance speedup of multiple GPUs against the same number of CPUs over a wide range of mesh sizes, demonstrating a maximum speedup of 25 times for 4--8 GPUs\slash CPUs. In addition, they found that for all but the largest problem sizes, MPI communication---rather than computation---became the performance scaling bottleneck.%

Xu et al.~\cite{Xu:2013gt} used GPUs to accelerate the expensive portions of a spectral-finite-difference DNS solver: the fast Fourier transform (FFT) and sparse linear equation solution. For the linear equation solver, they implemented both the standard GS and CG methods. While the standard GS is not appropriate for typical solution methods as discussed in Sec.~\ref{sec:laplace_meth}, the spectral-finite-difference approach results in a number of independent systems of linear equations, where only the equations corresponding to pairs of wavenumbers are interdependent. Xu et al.~\cite{Xu:2013gt} exploited this data independence for their GS implementation, using individual GPU threads to solve the linear equations for the wavenumber pairs. Their GPU implementation of the CG algorithm relied on NVIDIA's CUSPARSE library to accelerate sparse matrix-vector multiplications; similarly, their FFT implementation---combined with the GS and CG algorithms for the linear equation solution---used the CUFFT library for forward and backward FFTs. Xu et al.~\cite{Xu:2013gt} used both implementations to solve the scalar diffusion equation for problem sizes ranging over 64\textsuperscript{3}--240\textsuperscript{3}. The GS implementation performed up to 20 times faster than a serial CPU version, while the CG version demonstrated practically no speedup due to the small problem sizes involved in the sparse matrix operations. Xu et al.~\cite{Xu:2013gt} then applied the GS implementation to solving the three-dimensional Navier--Stokes equations, and demonstrated a performance speedup of nearly 26 times. However, due to memory limitations, they cautioned that their GS approach may not be suitable for larger problem sizes.%

\subsection{Reactive flow}
\label{sec:reactive}

In order to design the next-generation of engines and combustors, accurate and efficient simulations of reactive flow are vital. Traditionally, chemistry was represented in a simple manner, using one- or multiple-step global reaction mechanisms to capture the overall fuel oxidation and heat release. Unfortunately, the inability to capture pressure dependence, vital in high compression-ratio internal combustion engines, is one of the fundamental issues with such global mechanisms. In addition, as emissions regulations become more stringent, simulations must be able to predict concentrations of pollutant species and soot precursors for the development of advanced high-efficiency, low-emissions combustors.

Including detailed chemistry in simulations of reactive flow induces greater computational expense for two primary reasons: (1) chemical stiffness, caused by rapidly depleting species and\slash or fast reversible reactions, requires specialized integration algorithms (traditionally, high-order implicit solvers based on backward differentiation formulae); and (2) the large and ever-increasing size of detailed reaction mechanisms for transportation fuels of interest. While mechanisms for fuels relevant to hypersonic engines, such as hydrogen or ethylene, may contain 10--70 species~\cite{Burke:2011fh,Qin:2000ki}, a recent surrogate mechanism for gasoline consists of about 1550 species and 6000 reactions~\cite{Mehl:2011}; a mechanism for biodiesel surrogates contains almost 3300 species and over \num{10000} reactions~\cite{Herbinet:2010}. Incorporating such large, realistic reaction mechanisms in reactive flow simulations is beyond the scope of this survey; Lu and Law~\cite{Lu:2009gh} recently reviewed the subject. In brief, mechanism reduction is typically performed on large mechanisms to decrease the size from hundreds or thousands of species (and many more reactions) to a manageable size (e.g.,  $<$ 100 species); unfortunately, high-fidelity simulations using reaction mechanisms of this size are still too demanding to be performed on average computer systems. By utilizing the massively parallel architecture of relatively low-cost GPUs, previously inaccessible reactive-flow simulations may be performed on such systems. Though GPUs offer the potential of significant performance enhancement, researchers are only beginning to explore GPU acceleration of reactive-flow solvers.

Spafford et al.~\cite{Spafford:2010ky} made the first foray into this area by porting the species rate evaluation of Sandia National Laboratory's massively parallel DNS code S3D to the GPU. S3D is capable of solving the fully compressible Navier--Stokes equations combined with detailed chemistry~\cite{Hawkes:2005eh,Chen:2009gs}. The CPU performed the time integration of the chemical source terms with an explicit fourth-order Runge--Kutta method, where each integration step requires four species rate evaluations and consequently four GPU kernel invocations---and the associated memory transfer. In addition to calculating the rates for all grid points in parallel, the evaluation of reaction rates was accelerated by the GPU's hardware-accelerated transcendental function calls. Using an ethylene reaction mechanism with 22 species, they achieved a performance speedup of around 15 and 9 for single- and double-precision calculations, respectively.

In a different approach, Shi et al.~\cite{Shi:2011hv} used the GPU to simultaneously calculate all the reaction rates for a single kinetic system (i.e., a single computational volume\slash grid point) and accelerate the matrix inversion step of the implicit time integration using the GPU linear algebra library CULA~\cite{Humphrey:2010ga}. Initially, they demonstrated this using homogeneous autoignition simulations---essentially, the integration of a system of ordinary differential equations (ODEs) for temperature and the species rates. Overall, Shi et al.~\cite{Shi:2011hv} demonstrated a speedup of up to 22 for large chemical reaction mechanisms ($>$1000 species), but for moderate-size mechanisms ($<$100 species) the GPU version performed similarly or even slower than the CPU-only version.

Shi et al.~\cite{Shi:2011th} later implemented their GPU-based chemistry solver, paired with a dynamic mechanism reduction algorithm~\cite{Liang:2009,Shi:2010}, into the existing engine simulation code KIVA-Chemkin. Unlike the approach of Spafford et al.~\cite{Spafford:2010ky}, the species rates for each grid cells were evaluated in serial across the grid. In this case, acknowledging the poor performance of the GPU-based ODE solver on smaller mechanisms, the CPU performed the species integration when the mechanism size dropped below 300 species due to the dynamic reduction; the GPU-based solver only took over for mechanisms with $>$300 species. Using a large detailed mechanism for methyl decanoate (2877 species and 8555 reactions) to represent biodiesel~\cite{Herbinet:2008,Herbinet:2010}, Shi et al.~\cite{Shi:2011th} simulated a two-dimensional homogeneous-charge compression-ignition engine. The GPU-accelerated code performed nearly 20 times and twice as fast as the CPU-only version without and with the dynamic reduction, respectively. Therefore, if dynamic reduction is used, combining the GPU acceleration with this approach may not be needed since the mechanism size is sufficiently small during most of the simulation.

\begin{figure}[tbp]
\begin{center}
\includegraphics[width=.8\linewidth]{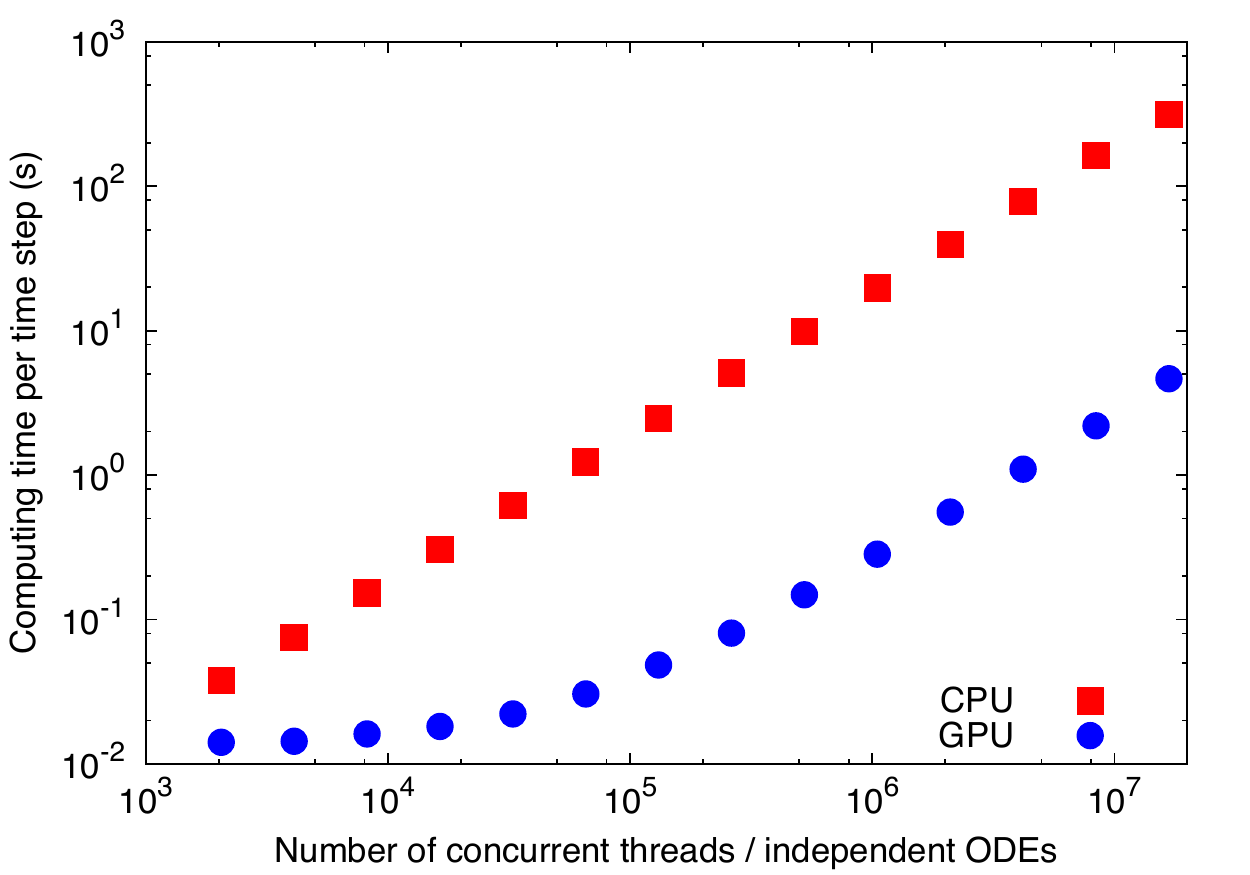}
\caption{Performance comparison of CPU and GPU explicit integration of a nonstiff hydrogen mechanism. Note that both axes are displayed in logarithmic scale.}
\label{fig:h2_gpu_explicit}
\end{center}
\end{figure}

Building upon the work of Spafford et al.~\cite{Spafford:2010ky}, Niemeyer et al.~\cite{Niemeyer:2011uw} developed a GPU-based explicit integration algorithm for nonstiff chemistry that integrated the species rate equations for all grid cells concurrently. In order to minimize memory transfer between the CPU and GPU, the entire time integration step of a fourth-order Runge--Kutta method was performed on the GPU. Contrast this with the approach of Spafford et al.~\cite{Spafford:2010ky}, where CPU-GPU communication must occur before and after each of the four species rate calculations. Using a compact hydrogen mechanism with 9 species and 38 irreversible reactions~\cite{Yetter:1991}, Niemeyer et al.~\cite{Niemeyer:2011uw} demonstrated a computational speedup of up to 75 compared to a single-core CPU over a wide range of independent spatial locations (i.e., cell volumes\slash grid points), as shown in Fig.~\ref{fig:h2_gpu_explicit}. The performance of the GPU over the CPU implementation improved with increasing number of computational cells, so this approach would be beneficial for large-scale simulations. Unlike the GPU-accelerated CFD approaches discussed above, the chemistry integration of the strategy of Niemeyer et al.~\cite{Niemeyer:2011uw} would be performed on the GPU simultaneously with CPU transport calculations, so no portion of the calculation waits for the other to finish.

Shi et al.~\cite{Shi:2012cl} presented a hybrid CPU\slash GPU chemistry integration strategy where, similar to the formulation of Niemeyer et al.~\cite{Niemeyer:2011uw}, the GPU explicitly integrated nonstiff chemistry in many grid cells concurrently. In their work, though, a standard implicit integrator on the CPU handled spatial locations with stiff chemistry. This combined approach achieved a performance speedup of 11--46. However, Shi et al.~\cite{Shi:2012cl} performed CPU transport calculations in serial with the CPU stiff chemistry integrations, limiting the potential performance.

Le et al.~\cite{Le:2013kt} developed the first reactive-flow solver that evaluated both the fluid transport and chemical kinetics terms on the GPU. As with most other approaches, they used operator splitting to decouple and independently solve the fluid transport and chemistry terms. They employed a first-order implicit method (the implicit\slash backward Euler method) to solve the stiff chemical kinetics terms, using a direct Gaussian elimination to solve the resulting linear system of equations. To implement their algorithms on the GPU, Le et al.~\cite{Le:2013kt} employed the same thread-per-cell approach as recent efforts~\cite{Spafford:2010ky,Niemeyer:2011uw,Shi:2012cl}. In the chemical kinetics kernel, Le et al.~\cite{Le:2013kt} attempted to use shared memory to store the system of variables as well as the Jacobian matrix (necessary for the Euler time integration), but could only store two rows of the matrix at a time due to the limited size of shared memory. Unfortunately, this led to excessive memory transfer between the global and shared memories, negatively impacting performance; using coalesced global memory alone offered better performance. Compared against an equivalent CPU version executed on a single processor core, their combined GPU solver performed up to 40 times faster using a reaction mechanism for methane with 36 species and 308 reactions, on a grid with over \num{e4} cells. However, the low accuracy of the chemistry solver---first order---should be noted.

Recently, Levesque et al.~\cite{Levesque:2012} presented their experiences hybridizing S3D from one level of parallelism using MPI to three levels, combining MPI (CPUs across multiple nodes), OpenMP (CPUs sharing memory on the same node), and OpenACC (GPUs on a node). Rather than developing native GPU code using CUDA or OpenCL, they used OpenACC paired with MPI and OpenMP to ensure architecture-agnostic code. To our knowledge, this was the first production CFD code ported to the GPU using OpenACC. Contrasted with the earlier work accelerating S3D~\cite{Spafford:2010ky}, Levesque et al.~\cite{Levesque:2012} moved all calculations to the GPUs, using the host CPUs to perform communication and boundary treatments.

In order to better parallelize the code on both OpenMP and OpenACC, Levesque et al.~\cite{Levesque:2012} made a number of changes to the structure of the code---although the actual algorithms and computations remained the same. For example, they separated the source term and derivative calculations, moved these terms into the same loop to maximizing spatial granularity and allowed overlapping of computation and communication. In another example, they developed a novel derivative queuing system that split the derivative computation into a number of tasks that could be overlapped with other calculations and communication. A number of other restructuring changes involved reduced memory operations and vectorized loops. In fact, optimizations targeted at GPU acceleration also resulted in improved CPU-only performance with OpenMP. The CPU-based MPI+OpenMP version ran up to around 1.4 times faster than the MPI version---using the same number of CPU cores. Levesque et al.~\cite{Levesque:2012} then compared the performance of the OpenMP and OpenMP+OpenACC codes on a single node and found that the GPU-accelerated version paired with a single 16-core CPU performed around 1.5 times faster than the OpenMP-only version on dual 16-core CPUs (32 total cores).

Stone et al.~\cite{Stone:2013jf} implemented two chemistry integrators on the GPU: (1) a fourth-order adaptive Runge--Kutta--Fehlberg (RKF45) method and (2) the standard fifth-order accurate implicit DVODE method. They applied these integrators to a reduced mechanism for ethylene with 19 species and 15 global reaction steps~\cite{Zambon:2007go}. Compared against equivalent single-core CPU versions over a range of ODE numbers, the RKF45 and DVODE methods on the GPU achieved speedup factors of up to 28.6 and 7.7, respectively. The lower relative performance increase of the implicit DVODE method is likely due to thread divergence induced by the complex logical flow inherent in the algorithm; the GPU-based DVODE method would demonstrate even less improvement compared against a parallelized CPU version operating on multiple cores (e.g., six). Stone et al.~\cite{Stone:2013jf} also compared the performance of the one-thread-per-ODE approach~\cite{Spafford:2010ky,Niemeyer:2011uw} against a new one-block-per-ODE approach, but found that the thread-based approach performs up to twice as fast for larger numbers of ODEs. At lower numbers of ODEs (e.g., $<$1000), the one-block-per-ODE approach is more efficient.

In a different manner from the direct evaluation of chemistry studied in prior efforts, Sankaran~\cite{Sankaran:2013he} presented a new approach for accelerating the chemistry in turbulent combustion modeling that used the GPU to solve the unsteady laminar flamelet equations. Flamelet models~\cite{Peters:1984td,Peters:1986wq} approximate the turbulent flame as a collection of embedded laminar flames. In the unsteady flamelet approach, rather than solving the equations for the species rates of change, certain controlling variables (e.g., mixture fraction, scalar dissipation rate) were used to determine the averaged\slash filtered species mass fractions through a probability density function (PDF) method. In Sankaran's~\cite{Sankaran:2013he} proposed approach, the CPU performed the main flow solver and tracked the distributions of mixture fraction and scalar dissipation rate. These values were then sent to the GPU, which integrated the unsteady flamelet equations explicitly. The resulting state variables were then integrated through a presumed PDF approach and returned to the host CPU. Sankaran's~\cite{Sankaran:2013he} method involved three levels of concurrency on the GPU: the solutions of (1) species reaction rates, thermochemical properties, and molecular transport rates; (2) the discretized flamelet equations in an regular grid in the mixture fraction space; and (3) multiple flamelets.

\subsection{Lattice Boltzmann method}
\label{sec:lbm}

The lattice Boltzmann method (LBM), reviewed by Chen and Doolen~\cite{Chen:1998co} and more recently by Aidun and Clausen~\cite{Aidun:2010ik}, developed as an alternative approach to simulating fluid flows. Rather than solving the Navier--Stokes equations, the macroscopic dynamics of the fluid are represented using a collection of microscopic particles. A simplified version of the full Boltzmann kinetic equation---the lattice Boltzmann equation---is then solved to determine the movement and collisions of each particle. Compared to more traditional Navier--Stokes-based solvers, LBM solvers allow easier treatment of complex flows and boundary conditions. In addition, the method is much less computationally expensive for high-Knudsen-number flows~\cite{Aidun:2010ik}. Since the core of the method relies on tracking a large number of independent particles, with only local movement and collision operations, it is well-suited to parallelism and shows highly scalable parallel performance~\cite{Aidun:2010ik}. As such, LBM is also suitable for operation on GPUs, and a number of researchers recently developed GPU-based implementations of such codes.

As with standard Navier--Stokes solvers, early efforts to implement the LBM on GPUs predated the development of programming environments such as CUDA and so required mapping computations to graphical operations such as that of the rasterization units and the frame buffer. Li et al.~\cite{Li:2003ke,Li:2005} developed (to our knowledge) the first implementation of the LBM on graphics hardware. With a grid size of 64\textsuperscript{3}, simulations ran fast enough to be performed in real time, while larger grids (128\textsuperscript{3}) remained interactive. Compared to a single-core CPU, the GPU version ran up to 15 times faster for single-precision computations. In another pre-CUDA work, Fan et al.~\cite{Fan:2004} developed an LBM solver for a cluster of GPUs, using it to study the dispersion of contaminants in Times Square. For a domain size of 480 $\times$ 400 $\times$ 80, their solver ran 4.6 times faster on a cluster of 30 GPUs than on the same number of CPUs. Zhao~\cite{Zhao:2007dr} followed the approach of Li et al.~\cite{Li:2003ke,Li:2005}, developing an LBM solver on the GPU to solve the diffusion, Laplace, and Poisson equations---although directed at modeling and visualization applications, such as the volume smoothing used in medical imaging techniques. Comparing with equivalent applications running on a single CPU core, Zhao~\cite{Zhao:2007dr} demonstrated a speedup factor of up to 184, although admittedly the CPU code was neither highly optimized nor parallelized.

Following the introduction of CUDA, a number of groups developed GPU-based LBM solvers, following implementation strategies similar to those found in the above Navier--Stokes-based solvers. T\"{o}lke and Krafczyk~\cite{Tolke:2008ig} developed an efficient solver and demonstrated it with simulations of a moving sphere in a pipe at a wide range of Reynolds numbers. In addition, they discussed optimization strategies for porting LBM codes to GPUs, resulting in a performance speedup of nearly two orders of magnitude. Bailey et al.~\cite{Bailey:2009hl} developed a GPU solver that ran over 28 times faster than an equivalent CPU version running on a quad-core CPU, via OpenMP. They achieved such high performance by (1) increasing GPU multiprocessor occupancy, (2) cutting memory requirements in half through a new memory access technique, and (3) exploiting global memory coalescing by judicious use of shared memory.

Bernaschi et al.~\cite{Bernaschi:2010ja} accelerated the LBM portion of MURPHY, a multi-scale simulation code for fluids with embedded particles that combines LBM to capture fluid flow with a modified molecular dynamics solver for suspended solid particles. Unlike pure LBM solvers, in this coupled code some information must be transferred back to the CPU during the simulation for the molecular dynamics portion. Data representing the large population of fluid particles remained on the GPU, while a smaller number of hydrodynamic variables were transferred each time step. Compared to a CPU version running on a quad-core processor, the LBM portion ran eight to ten times faster on a GPU. In a later work, Bernaschi et al.~\cite{Bernaschi:2012ei} ported the molecular dynamics portion of MURPHY such that the code ran fully on the GPU, which further improved performance by a factor of five to six over the prior GPU version.

Kuznik et al.~\cite{Kuznik:2010eo} used a GPU-based LBM solver to simulate flow in a lid-driven cavity, showing that single-precision results were nearly indistinguishable---as well as being almost four times faster---than double-precision calculations. Obrecht et al.~\cite{Obrecht:2011} also used the lid-driven cavity case to demonstrate techniques for optimizing the LBM on GPUs, reaching 86\% of the theoretical global memory throughput for their GPU card.

Rinaldi et al.~\cite{Rinaldi:2012jm} developed a number of novel techniques for implementing LBM solvers on GPUs, in addition to benefitting from lessened restrictions on memory access in a new CUDA version. In order to minimize data transfer, the particle collision and propagation steps were merged into a single loop. In addition, the entire algorithm was performed in fast shared memory before being written to global memory to return to the host CPU. In simulations of the familiar three-dimensional lid-driven cavity problem, this optimized solver performed 92--130 times faster than a CPU version running on a single core, for single-precision calculations.

\subsection{Summary of findings}
\label{sec:summary}

\begin{table}[htbp]
\begin{center}
\begin{tabular}{@{}l l l l@{}}
\toprule
Topic & Technique & Improvement & References \\
\midrule
Memory & Coalesced global memory & 1.4--1.7$\times$ & \cite{Tolke:2008ig,Bailey:2009hl,Cohen:2009tp,Asouti:2010gy,Griebel:2010kb,Kampolis:2010jf,Phillips:2010tn,Corrigan:2011ig,Gohari:2012bd,KhajehSaeed:2013kz,Le:2013kt,Salvadore:2012} \\
 & Avoiding local memory & - & \cite{Hagen:2006uq,Griebel:2010kb,Salvadore:2012,Zaspel:2012gp} \\
 & Shared memory & 1.8--2.2$\times$ & \cite{Brandvik:2007bn,Brandvik:2008up,Tolke:2008ig,Bailey:2009hl,Phillips:2009vj,Asouti:2010gy,Kampolis:2010jf,Alfonsi:2011kn,Ran:2011,Lefebvre:2012,Thibault:2012kg,KhajehSaeed:2013kz,Salvadore:2012,Zaspel:2012gp} \\
 & Texture memory & 1.1--1.4$\times$ & \cite{Scheidegger:2005we,Hagen:2006uq,Brandvik:2007bn,Brandvik:2008up,Elsen:2008jv,Cohen:2009tp,Asouti:2010gy,Kampolis:2010jf,Shinn:2010wf,Kuo:2011jo,Vanka:2011vc} \\
 & Constant memory & - & \cite{Asouti:2010gy,Kampolis:2010jf,Alfonsi:2011kn,KhajehSaeed:2013kz} \\
 & \multirow{2}{3cm}{Overlap memory access\slash computation} & 1.1--3.9$\times$ & \cite{Corrigan:2011ig,Lefebvre:2012} \\ \\ \midrule
Communication & \multirow{2}{3cm}{Asynchronous communication} & 1.4$\times$ & \cite{Hagen:2006uq,Griebel:2010kb,Phillips:2010tn,Lefebvre:2012,Zaspel:2012gp,Levesque:2012,Bernaschi:2012ei,DeLeon:2013vp,Jacobsen:2013ik,KhajehSaeed:2013kz} \\ \\
 & Move entire code to GPU & 1.3--5$\times$ & \cite{Brandvik:2007bn,Brandvik:2008up,Elsen:2008jv,Phillips:2010tn,Brandvik:2011dt,Ran:2011,Lefebvre:2012,Levesque:2012,Bernaschi:2012ei,Gohari:2012bd,Salvadore:2012,Zaspel:2012gp} \\
 & Pinned memory & 3$\times$ & \cite{Lefebvre:2012,Zaspel:2012gp} \\ \midrule
\multirow{2}{1.5cm}{Boundary treatment} & \multirow{2}{3cm}{Group all boundary\slash ghost nodes} & 2$\times$ & \cite{Phillips:2009vj} \\ \\ \midrule
Precision & Single over double & 1.4--4.5$\times$ & \cite{Kuznik:2010eo,Kampolis:2010jf,Spafford:2010ky,Corrigan:2011ig,Thibault:2012kg} \\
 & Mixed over double & 1.3--1.4$\times$ & \cite{Kampolis:2010jf} \\
\midrule
Thread workload & One block\slash ODE & 0.8--2$\times$ & \cite{Stone:2013jf} \\
\bottomrule
\end{tabular}
\caption[Summary of techniques to improve performance in GPU-based CFD codes.]{Summary of techniques to improve performance in GPU-based CFD codes. The factor of improvement refers to the value reported for the specific technique employed, rather than the speedup of the entire GPU code. A ``-'' indicates no value was reported.}
\label{tab:summary}
\end{center}
\end{table}

Table~\ref{tab:summary} summarizes our findings for specific optimization techniques that have been employed in GPU-accelerated CFD applications. Specific improvement factors for each technique are included where reported. Interested readers should peruse the listed references for specific implementation details, although many were described in general here.

Regarding precision, CUDA-capable GPUs with a compute capability of at least 1.3 (since 2007) mostly support the IEEE 754 standard~\cite{IEEE:2008}, with a few deviations. When first introduced, double-precision operations suffered a severe performance decrement compared to single precision, but recent GPU hardware~\cite{NVIDIA:2010a} offers nearly the same relative double-to-single precision performance as CPUs. Certain applications, such as solving the Euler equations~\cite{Hagen:2006uq,Elsen:2008jv}, do not require double precision, but others may need the higher precision. Elsen et al.~\cite{Elsen:2008jv} reported that, for single-precision calculations, GPU results compared closely with those from the CPU up to five or six digits. Judicious choice of precision, contrasting specific needs with desired performance, is warranted based on the application.

\section{Concluding remarks}
\label{sec:conclusion}

In this article, we reviewed the progress made in developing GPU-based computational fluid dynamics solvers, covering efforts capable of simulating laminar, turbulent, and reactive flows, in addition to solvers based on the lattice Boltzmann method. Initial efforts took existing applications and transferred some or most of the calculations to the GPU for acceleration, demonstrating promising speedup of computations. Later works used the GPU to perform nearly all calculations, leading to improved performance by minimizing data transfer. In addition, using case studies of heat conduction in a plate and lid-driven cavity flow, we demonstrated a systematic manner to optimize GPU codes. These case studies also served to show the importance of redesigning numerical algorithms for execution on GPUs, even for simple applications, as well provide examples of potential performance increase. In general, we observed trends of around an order of magnitude improvement in performance compared to versions running on multicore CPUs.

In a number of early efforts, comparisons were made between the performance of a GPU code and a serial CPU code running on a single processing core. Since nearly all modern CPUs consist of at least two---and commonly four or more---cores, and parallelization via OpenMP or MPI is common in scientific computing, future comparisons made should include a CPU version running in parallel on multiple cores. In addition, comparing the performance normalized by both hardware cost and power consumption, as suggested by Zaspel and Griebel~\cite{Zaspel:2012gp}, provides even more insight to the benefits of moving computationally expensive CFD codes to graphics processors.

Through analyzing the patterns of GPU implementations surveyed here as well as the demonstrated case studies, a number of successful strategies emerged:
\begin{itemize}
	\item Global memory should be arranged to coalesce read/write requests, which can improve performance by an order of magnitude (theoretically, up to 32 times: the number of threads in a warp).
	\item Shared memory should be used for global reduction operations (e.g., summing up residual values, finding maximum values) such that only one value per block needs to be returned.
	\item Use asynchronous memory transfer, as shown by Phillips et al.~\cite{Phillips:2010tn} and DeLeon et al.~\cite{DeLeon:2013vp} when parallelizing solvers across multiple GPUs, to limit the idle time of either the CPU or GPU.
	\item Minimize slow CPU-GPU communication during a simulation by performing all possible calculations on the GPU.
\end{itemize}

A number of challenges remain in developing GPU solvers for simulations of flows with more complex physics. For reactive-flow calculations, most strategies presented so far~\cite{Spafford:2010ky,Niemeyer:2011uw,Shi:2011hv,Shi:2011th,Shi:2012cl} performed the integration of the chemistry terms on the GPU while the controlling CPU handled transport calculations. Solvers based on this approach necessarily perform repeated CPU-GPU memory transfer of the thermochemical and species mass fraction variables, and suffer from the corresponding added latency. This could be balanced by the performance increase of calculating these expensive terms on the GPU, but it should be investigated whether migrating all calculations to the GPU would result in even higher performance, at the risk of leaving the CPU unnecessarily idle. Furthermore, no practical demonstration of a GPU chemistry solver capable of handling stiff chemistry has yet been made. This is one area where efforts need to be focused.

Only two studies, from the same group, have shown a GPU-based solver capable of simulating of two-phase, incompressible flows using level-set methods~\cite{Griebel:2010kb,Zaspel:2012gp}, related to a rising air bubble in water. However, other than these efforts, no GPU-accelerated simulations with other complex physical models such as multiphase flows have yet been demonstrated, even though such phenomena are important in a number of applications. Physical models that rely on Lagrangian schemes (e.g., for sprays or particles) in particular stand to benefit from GPU acceleration due to the data independence of tracking the movement of notional particles or droplets.

A common question regarding GPU computing may be, ``How much can an application benefit from GPU acceleration?'' Amdahl's law~\cite{Amdahl:1967} gives the answer, stating that the overall speedup achievable $N$, if a proportion $P$ of the code can be accelerated by a factor of $S$, is
\begin{equation}
N = \frac{1}{ \left( 1 - P \right) + \frac{P}{S} } .
\end{equation}
Using this equation, if 50\% of the application can be GPU-accelerated then the overall speedup is $2S/(1+S)$. Even if the portion performed on the GPU can be accelerated by a factor of 100, the overall application will only perform about twice as fast. On the other hand, if 99\% of the application is parallelizable, then the overall potential speedup is around $100S/(99+S)$. Taking the case of Niemeyer et al.~\cite{Niemeyer:2011uw} as an example, for larger grid sizes the evaluation of the reactive source terms could be accelerated by a factor of 75. In this case, the overall speedup could reach about 43 times. Therefore, taking full advantage of GPU acceleration can only be achieved if most of the application is parallelizable.

Early GPU work in many areas tackled ``low-hanging fruit'' problems that required little to no significant change in program structure. For example, an application that involves frequent large matrix inversions could simply replace the subroutine call from the CPU version to the GPU version. However, most software was designed to run on traditional, sequential processors (e.g., CPUs), so this ``incremental'' approach can only help so far. In order to truly harness the massive parallelism of many-core processors (and GPUs in particular), new programming strategies and new algorithms must be designed specifically targeted at massively parallel processors. In particular, reactive-flow solvers that require complex, implicit integration methods for stiff chemistry at every spatial location cannot be simply ported to the GPU, so alternative approaches must be developed in order to achieve a more useful improvement in performance. By developing new solvers for fluid flow simulations using GPUs to significantly improve performance, we can facilitate the transition of high-fidelity CFD from a tool of science to that of technology and design, enabling previously inaccessible levels of detail and parametric variation in modeling studies.

\subsection*{Acknowledgments}

This work was supported by the National Science Foundation under grant number 0932559, the US Department of Defense through the National Defense Science and Engineering Graduate Fellowship program, the National Science Foundation Graduate Research Fellowship under grant number DGE-0951783, and the Combustion Energy Frontier Research Center---an Energy Frontier Research Center funded by the US Department of Energy, Office of Science, Office of Basic Energy Sciences under award number DE-SC0001198.


\end{document}